Banner appropriate to article type will appear here in typeset article

# Force balances in spherical shell rotating convection

S. Naskar[1]†, C. J. Davies[1], J. E. Mound[1], A. T. Clarke[1]

[1]School of Earth and Environment, University of Leeds, LS2 9JT, Leeds, UK

Significant progress has been made in understanding planetary core dynamics using numerical models of rotating convection (RC) in spherical shell geometry. However, the behaviour of forces in these models within various dynamic regimes of RC remains largely unknown. Directional anisotropy, scale dependence, and the role of dynamically irrelevant gradient contributions in incompressible flows complicate the representation of dynamical balances in spherical shell RC. In this study, we systematically compare integrated and scale-dependent representations of mean and fluctuation forces and curled forces (which contain no gradient contributions) separately for the three components $(\hat{r}, \hat{\theta}, \hat{\phi})$. The analysis is performed with simulations in a range of convective supercriticality $Ra/Ra_c = 1.2 - 1967$ and Ekman number $E = 10^{-3} - 10^{-6}$, with fixed Prandtl number $Pr = 1$, no-slip and fixed flux boundaries. We exclude 10 velocity boundary layers from each boundary of the spherical shell, which provides a consistent representation of the dynamics between the volume-averaged force and curled force balance in the parameter space studied. Radial, azimuthal and co-latitudinal components exhibit distinct force and curled force balances. The total magnitudes of the mean forces and mean curled forces exhibit a primary thermal wind (TW) balance; the corresponding fluctuating forces are in a quasi-geostrophic (QG) primary balance, while the fluctuating curled forces transition from a Viscous-Archimedean-Coriolis balance to an Inertia-Viscous-Archimedean-Coriolis balance with increasing $Ra/Ra_c$. The curled force balances are more weakly scale-dependent compared to the forces and do not show clear cross-over length scales. The fluctuating force and curled force balances are broadly consistent with three regimes of RC (weakly nonlinear, rapidly rotating, and weakly rotating), but do not exhibit sharp changes with $Ra/Ra_c$, which inhibits the identification of precise regime boundaries from these balances.

## 1. Introduction

Buoyancy-driven convection in a rotating spherical shell is a classical framework for studying the dynamics and magnetic field generation in the cores of planets and stars. However, despite significant progress (Gastine *et al.* 2016; Schaeffer *et al.* 2017; Long *et al.* 2020; Schwaiger *et al.* 2020; Gastine & Aurnou 2023), state-of-the-art direct numerical simulations cannot reach the parameter values representative of the astrophysical bodies and are unlikely to do so in the near future (Davies *et al.* 2011; Roberts & King 2013). Therefore, extensive work has focused on developing scaling relations between the governing input parameters and global output diagnostic quantities of numerical simulations, which can be used to extrapolate to the conditions of planetary cores and facilitate comparisons with available observations. These scaling relations rely on the balance of forces that determine the system dynamics (e.g. Christensen *et al.* 2010; King & Buffett 2013; Aubert *et al.* 2017), and so it is crucial to quantify dynamical balances in numerical simulations accurately.

† Email address for correspondence: S.Naskar@leeds.ac.uk

The dynamics in rotating non-magnetic convection are governed by the Ekman number $E$, the ratio of viscous to Coriolis forces, the Prandtl number $Pr$, the ratio of viscous to thermal diffusivities, and the Rayleigh number $Ra$, measuring the buoyancy force driving convection. Theoretical considerations suggest that the convecting system exhibits at least three distinct dynamical regimes (e.g. Aubert *et al.* 2001; King & Buffett 2013; King *et al.* 2013; Gastine *et al.* 2016; Long *et al.* 2020; Aurnou *et al.* 2020; Kunnen 2021). At a fixed $E$ and $Pr$, raising the thermal forcing to just above the onset of convection (i.e, $Ra \geqslant Ra_c$, where $Ra_c \propto E^{4/3}$ is the critical Rayleigh number at the onset of convection) leads to weakly-nonlinear (WN) convection. In this regime, the primary dynamics is expected to be a quasi-geostrophic (QG) balance between Coriolis and pressure forces, while the residual (i.e., the ageostrophic Coriolis force) is balanced by buoyancy and viscous forces forming a secondary Viscous-Archimedean(buoyancy)-Coriolis(ageostrophic) or VAC balance. Increasing the thermal forcing increases the role of inertia, which is expected to lead to a weakly rotating (WR) regime where the primary QG balance is gradually broken, eventually resulting in non-rotating behaviour at sufficiently high $Ra$. Between the WN and WR regimes, a rapidly rotating (RR) regime is expected for a range of thermal forcing, where both viscosity and inertial forces are small relative to the Coriolis force in the primary QG balance. Therefore, the RR regime is thought to be more relevant for investigating planetary core convection, as compared to the other regimes (Kunnen 2021).

Depending on the dynamical balance, global and local flow diagnostic quantities (e.g. average velocity, length scale, heat transport, and boundary layer thickness) may exhibit distinct scaling with the governing input parameters. Scaling regimes in rotating spherical shell convection have been extensively studied by Gastine *et al.* (2016) and Long *et al.* (2020) (henceforth referred to as *L*20). The conformity of the simulation diagnostics with theoretical scaling laws was utilized to demarcate boundaries between the dynamical regimes (i.e., WN, RR, WR). For example, Gastine *et al.* (2016) and *L*20 defined the WN regime based on the theoretical expectation that at low supercriticality, the Nusselt number $Nu$, representing the ratio of the total average heat flux from the shell to the conductive flux, scales as $Nu - 1 \propto Ra/Ra_c - 1$ (Gillet & Jones 2006). They found that this heat transfer behaviour is consistent with the dimensionless flow lengthscale scaling $\ell \sim E^{1/3}$ and Reynolds number scaling $Re \sim B^{1/2}E^{1/3}$ expected in the VAC regime, where $B$ is the convective power. These comparisons relied on assumed force balances; however, the quantitative behaviour of forces in numerical simulations of spherical shell RC remains largely unexplored. A few studies have explored force balances in plane layer geometries (Guzmán *et al.* 2020; Naskar & Pal 2022*a*,*b*), which may represent the dynamics in the tangent cylinder region of a spherical shell (Gastine & Aurnou 2023), while Schwaiger *et al.* (2020), Teed & Dormy (2023) and Nicoski *et al.* (2024) have computed force balances in a limited number of spherical shell RC runs. Hence, the primary objective of our study is to investigate the dynamical balances that emerge in numerical simulations of spherical shell RC.

Quantifying dynamical balances in rotating convection is an intricate issue. The force representation in incompressible flows is complicated by the existence of the gradient portions of forces, which are balanced by the pressure gradient term, but are not directly relevant to the dynamics (Hughes & Cattaneo 2019). A simple way to remove the gradient contributions is to curl the force balance (Dormy 2016), though the derivative operation can enhance small-scale contributions to the individual terms (Teed & Dormy 2023). Differences between the dynamics predicted by force and curled balances are important since some theoretical predictions consider the asymptotic behaviour of forces (Nicoski *et al.* 2024), whereas others are based on curled forces (e.g. VAC and IAC in *L*20). Teed & Dormy (2023) found that forces and their curls predicted different dynamical balances in a single simulation of rotating spherical shell convection. Here we investigate the consistency between force and curled depictions of rotating convection dynamics across a broad range of parameters ($Ra/Ra_c = 1.1 - 300$, $E = 10^{-3} - 10^{-6}$ and $Pr = 1$).

When calculating forces or curls, it is crucial to distinguish dynamics in the boundary layers and the convective bulk. Integrating forces over the entire spherical shell may overestimate the role of viscosity in the bulk dynamics (Soderlund *et al.* 2012; Yadav *et al.* 2016). Most studies of force calculations remove a region corresponding to one velocity boundary layer (VBL) thickness at the top and bottom of the domain (e.g. Schwaiger *et al.* 2020; Teed & Dormy 2023; Nicoski *et al.* 2024) in order to isolate the bulk dynamics. However, we are unaware of previous systematic studies that have established the thickness of the layer near the boundaries that should be excluded from the volume-averaged forces/curls to obtain a robust estimate of the bulk dynamics.

In rotating spherical shell convection, the axisymmetric part of forces, corresponding to spherical harmonic order $m = 0$, does not contribute to the convective dynamics. Nicoski *et al.* (2024) partitioned the forces into azimuthally averaged and corresponding fluctuating parts and analysed the scaling behaviour of the fluctuating radial forces as a function of $E$ and $Ra$ in RC with strong zonal flows, finding a primary QG balance across a broad range of parameters. In contrast to the QG force balance in the small-scale (i.e., $m \neq 0$) convective motions (Nicoski *et al.* 2024), Aubert (2005) focused on the large-scale dynamics and used curls to identify a horizontal thermal wind (TW) balance in the large-scale azimuthal ($m = 0$) motions for RC simulations operating in the RR regime. Here, we systematically compare mean and fluctuating components of forces and their curls across different dynamical regimes of spherical shell RC.

In a spherical shell geometry, an additional difficulty in representing the system dynamics arises due to the dependence of each force and curl component ($\hat{r}, \hat{\theta}, \hat{\phi}$) on the spatio-temporal co-ordinates ($r, \theta, \phi, t$) and scale (spherical harmonic degree $l$ and order $m$), as well as the governing parameters ($Ra, E, Pr$). Since the buoyancy force is radial while the system rotates about a vertical axis, different dynamical balances may be expected in the three orthogonal directions. Directional anisotropy of the force balance has been rarely considered in spherical shell models (see Calkins *et al.* (2021) for an exception), and forces are usually represented by their total magnitudes (Guzmán *et al.* 2021; Orvedahl *et al.* 2021; Soderlund *et al.* 2012). Studies of azimuthally averaged forces (Calkins *et al.* 2021) and curls (Aubert 2005) in dynamo simulations have indeed revealed a TW balance in the meridional plane with a Coriolis-Lorentz balance in the azimuthal direction. We complement these works by considering the balance of individual ($\hat{r}, \hat{\theta}, \hat{\phi}$) force components in rotating spherical shell convection.

The scale-dependence of the force balance in RC spherical shell simulations has been investigated by considering the force contributions from each spherical harmonic degree $l$. Schwaiger *et al.* (2020) found a secondary balance between ageostrophic Coriolis and buoyancy force at low $l$ and a cross-over to Coriolis-Inertia balance at higher $l$. They argued that the length scale at which inertia and buoyancy forces cross over is related to energetic scales of the flow, as estimated from the peak of the poloidal kinetic energy spectra. However, the scale-dependent representation of curled forces may not exhibit such a dynamically relevant cross-over scale (Teed & Dormy 2023). Furthermore, some simulations show that the small-scale curled balance differs markedly compared to the force balance (Teed & Dormy 2023). It is, therefore, useful to systematically compare the scale-dependent nature of force and curled balances in different dynamical regimes of RC.

In summary, previous studies have investigated various aspects of dynamical balances (e.g., forces vs. curls, mean vs. fluctuating balances, components vs. combined terms, scale-dependence) in spherical shell RC. However, to our knowledge, no study systematically compares the different depictions of dynamical balances in different regimes of spherical shell RC. A recent related study by Nicoski *et al.* (2024) investigated the behaviour of fluctuating radial forces in spherical shell RC with fixed temperature and free-slip boundaries for $E = 5 \times 10^{-4} - 5 \times 10^{-7}$, $Pr = 1$ and the convective supercriticality range $3 - 161$. They found a primary QG balance in the radial fluctuating forces at all parameters considered. At low convective supercriticalities, they obtained a secondary balance between Buoyancy and ageostrophic Coriolis, while inertia and viscosity enter the balance at higher thermal forcing. In this paper, we systematically compare integrated and scale-dependent representations of mean and fluctuating force and curled balances in the large suite of rotating spherical shell convection simulations from Mound & Davies (2017) and *L*20, supplemented by three new simulations. In contrast to Nicoski *et al.* (2024), our simulations use no-slip, fixed heat flux boundaries.

This paper is organised as follows. We present the mathematical model in section 2.1, the numerical details in section 2.2, and the force calculation method in section 2.3. Results are presented in Section 3, beginning in Section 3.1 with an analysis of the boundary region that must be excluded to ensure a robust representation of the bulk dynamics. In sections 3.2 and 3.3, we investigate mean and fluctuating bulk-integrated force and curl balances, respectively. The total magnitude of the forces and their scale dependence is discussed in 3.4. In Section 4, we compare transitions in the computed force and curled balanced to the regime diagram obtained by *L*20. We discuss our observations and summarize the findings in section 5.

## 2. Method

### 2.1. *Mathematical model*

We employ a numerical model of convection of a Boussinesq fluid in a rotating spherical shell. The relevant physical properties of the fluid are the kinematic viscosity, $\nu$, thermal expansivity, $\alpha$, and thermal diffusivity, $\kappa$, defined as $\kappa = k/\rho_0 c_p$, where $k$ is the thermal conductivity, $\rho_0$ is the reference density, and $c_p$ is the specific heat capacity. A spherical coordinate system $(r, \theta, \phi)$ is used to represent the domain bounded by the inner and outer boundaries, $r_i$ and $r_o$, respectively. The system rotates with a constant angular velocity $\boldsymbol{\Omega} = \Omega\hat{z}$, about the vertical and gravity $g$ varies linearly with radius, with $g = g_o$ at the outer radius. The governing equations are cast in non-dimensional form using the shell gap $h$ as the length scale, the viscous diffusion time, $h^2/\nu$, as the time scale, and $\beta/h$ as the temperature scale, giving

$$\boldsymbol{\nabla} \cdot \boldsymbol{u} = 0 \tag{2.1}$$

$$\frac{\partial \boldsymbol{u}}{\partial t} + (\boldsymbol{u} \cdot \boldsymbol{\nabla})\boldsymbol{u} + \frac{1}{E}(\hat{z} \times \boldsymbol{u}) = -\boldsymbol{\nabla}\widetilde{P} + \left(\frac{Ra}{Pr}\right)Tr + \nabla^2\boldsymbol{u} \tag{2.2}$$

$$\frac{\partial T}{\partial t} + (\boldsymbol{u} \cdot \boldsymbol{\nabla})(T + T_c) = \frac{1}{Pr}\nabla^2(T + T_c) \tag{2.3}$$

where $T$ is the temperature fluctuation relative to the conductive state $T_c$, given by $\partial T_c/\partial r = -\beta/r^2$. The parameter $\beta$ is related to the fixed heat flow through the boundaries as $Q = 4\pi\beta k$. No-slip velocity conditions and fixed heat flux temperature conditions have been used at both boundaries.

The non-dimensional numbers appearing in these equations are the Ekman number ($E$), Rayleigh number ($Ra$), and Prandtl number ($Pr$), which are defined as,

$$E = \frac{\nu}{2\Omega h^2}, \qquad Ra = \frac{g_o \alpha \beta h^3}{\nu \kappa r_o}, \qquad Pr = \frac{\nu}{\kappa}. \tag{2.4}$$

### 2.2. *Numerical details*

The velocity field is represented by toroidal and poloidal scalar fields, which are expressed as radially varying Schmidt-normalized spherical harmonics. Radial variations are expressed using second-order finite differences on the zeros of Chebyshev polynomials. A predictor-corrector scheme is used for time stepping in spectral space that treats the diffusion terms implicitly. Further numerical details can be found in previous studies that use the same solver (Willis *et al.* 2007; Davies *et al.* 2011; Matsui *et al.* 2016).

We consider all the simulations reported in Mound & Davies (2017) for $E = 10^{-4}, 10^{-5}$ and $10^{-6}$, that impose homogeneous heat flux at the outer boundary. The simulations performed by $L20$ at $E = 10^{-3}$, $E = 3 \times 10^{-4}$ and $E = 3 \times 10^{-5}$ are also included. It should be noted here that the flux Rayleigh number in this study relates to the modified flux Rayleigh number defined by Mound & Davies (2017) and $L20$, as $\widetilde{Ra} = RaE$. To complement the database, we have run three more simulations at $E = 10^{-6}$, for $\widetilde{Ra} = 350, 550$ and $30000$. The details of these simulations are given in appendix A. The Prandtl number is fixed at $Pr = 1$ for all runs. The fixed temperature Rayleigh number ($Ra_T$) is related to the fixed flux Rayleigh number ($Ra$) as,

$$Ra_T = \frac{Ra}{Nu} \frac{(1-\eta)^2}{\eta} \tag{2.5}$$

where $\eta = r_i/r_o$ is the radius ratio fixed at $\eta = 0.35$. The reduced fixed-temperature Rayleigh number used in section 3 is defined as $\widetilde{Ra_T} = Ra_T E^{4/3}$.

### 2.3. *Force calculation*

We refer to the terms from left to right in equation 2.2 as time-derivative ($TD$), inertia ($I$), Coriolis ($C$), pressure gradient ($P$), thermal buoyancy (or Archimedean, $A$) and the viscous ($V$) forces. Additionally, the ageostrophic Coriolis force $\frac{1}{E}(\hat{z} \times \boldsymbol{u}) + \boldsymbol{\nabla}\widetilde{P}$ is denoted as $C_{ag}$. Henceforth, we use these abbreviations to refer to the balance in the simulations. For example, a balance between Inertia, Archimedean (i.e., thermal buoyancy) and Coriolis forces will be referred to as an IAC balance. Also, a TW balance refers to an ACP force balance or an AC balance of curled forces. Similarly, a QG balance refers to a CP balance of forces.

We partition dependent variables into their azimuthally averaged mean and corresponding fluctuating parts as,

$$\begin{aligned} f(r,\theta,\phi,t) &= \overline{f}(r,\theta,t) + f'(r,\theta,\phi,t) \\ \overline{f}(r,\theta,t) &= \frac{1}{2\pi} \int_0^{2\pi} f(r,\theta,\phi,t) d\phi \end{aligned} \tag{2.6}$$

In spectral representation, this is equivalent to partitioning into the spherically symmetric harmonics of order $m = 0$ (mean) and the corresponding part with order $m \neq 0$. Azimuthally averaging equation 2.2 leads to

$$\underbrace{\frac{\partial \overline{\boldsymbol{u}}}{\partial t}}_{\overline{TD}} + \underbrace{(\overline{\boldsymbol{u}} \cdot \boldsymbol{\nabla})\overline{\boldsymbol{u}}}_{I^{mm}} + \underbrace{\overline{(\boldsymbol{u}' \cdot \boldsymbol{\nabla})\boldsymbol{u}'}}_{\overline{I^{ff}}} + \underbrace{\frac{1}{E}(\hat{z} \times \overline{\boldsymbol{u}})}_{\overline{C}} = \underbrace{-\boldsymbol{\nabla}\overline{P}}_{\overline{P}} + \underbrace{\frac{Ra}{Pr}\overline{T}\,\boldsymbol{r}}_{\overline{A}} + \underbrace{\nabla^2 \overline{\boldsymbol{u}}}_{\overline{V}} \qquad (2.7)$$

where partitioning the mean inertial force term $\overline{I} = \overline{(\boldsymbol{u} \cdot \boldsymbol{\nabla})\boldsymbol{u}}$, as $\overline{I} = I^{mm} + \overline{I^{ff}}$, leads to the mean-mean inertia term $I^{mm} = (\overline{\boldsymbol{u}} \cdot \boldsymbol{\nabla})\overline{\boldsymbol{u}}$ and the Reynolds stress $\overline{I^{ff}} = \overline{(\boldsymbol{u}' \cdot \boldsymbol{\nabla})\boldsymbol{u}'}$. Subtracting the mean momentum equation 2.7 from equation 2.2 leads to the corresponding fluctuating part of the momentum equation,

$$\underbrace{\frac{\partial \boldsymbol{u}'}{\partial t}}_{TD'} + \underbrace{(\overline{\boldsymbol{u}} \cdot \boldsymbol{\nabla})\boldsymbol{u}'}_{I^{mf}} + \underbrace{(\boldsymbol{u}' \cdot \boldsymbol{\nabla})\overline{\boldsymbol{u}}}_{I^{fm}} + \underbrace{(\boldsymbol{u}' \cdot \boldsymbol{\nabla})\boldsymbol{u}'}_{I^{ff}} - \underbrace{\overline{(\boldsymbol{u}' \cdot \boldsymbol{\nabla})\boldsymbol{u}'}}_{\overline{I^{ff}}} + \underbrace{\frac{1}{E}(\hat{z} \times \boldsymbol{u}')}_{C'} =$$
$$\underbrace{-\boldsymbol{\nabla} P'}_{P'} + \underbrace{\frac{Ra}{Pr}T'\,\boldsymbol{r}}_{A'} + \underbrace{\nabla^2 \boldsymbol{u}'}_{V'} \qquad (2.8)$$

where the fluctuating inertial term $I'$ has four parts $I' = I^{mf} + I^{fm} + I^{ff} - \overline{I^{ff}}$, signifying the mean-fluctuating, fluctuating-mean, fluctuating-fluctuating inertial terms and the Reynolds stress. Among the fluctuating inertial terms in 2.8, the fluctuating-fluctuating term ($I^{ff}$) always dominates in our simulations. Therefore, we have only discussed the sum of the four terms $I'$, for presentational convenience. In our calculations, the temporally-averaged values of time derivative terms (TD) in equations 2.7 and 2.8 remain small compared to all the other terms, though it may be significant for simulations with free-slip boundaries (Nicoski *et al.* 2024). Therefore, this term is not discussed in the next section and the non-linear advection terms are referred to as inertia throughout the text. Also, the abbreviations of mean (fluctuating) forces are mentioned explicitly rather than using overbar (prime). For example, a Viscous-Archimedean-Coriolis balance in the fluctuating curled forces will be referred to as a curled fluctuating $VAC$ balance, rather than a $V'A'C'$ balance of curled forces.

The total force magnitude can be calculated from the vector components as,

$$f_{tot} = |\boldsymbol{f}| = \sqrt{f_r^2 + f_\theta^2 + f_\phi^2} \qquad (2.9)$$

where $f_r$, $f_\theta$, and $f_\phi$ are the components in $\hat{r}$, $\hat{\theta}$, and $\hat{\phi}$ directions respectively. Further, the force components (or the total force magnitude) are represented by their root mean square (r.m.s.) values, where the "mean" refers to a volume average performed by first averaging over a spherical surface and then averaging in the radial direction excluding regions near the boundary,

$$\langle f_j \rangle_S = \frac{1}{4\pi r^2} \int_0^\pi \int_0^{2\pi} f_j(r, \theta, \phi, t)\, r^2 \sin\theta\, d\phi\, d\theta, \qquad (2.10a)$$

$$\langle f_j \rangle_V = \frac{1}{h - r_{ex}\delta_v^i - r_{ex}\delta_v^o} \int_{r_i + r_{ex}\delta_v^i}^{r_o - r_{ex}\delta_v^o} \langle f_j \rangle_S\, dr, \qquad (2.10b)$$

where $j = r, \theta, \phi$ or $tot$. In equation 2.10b, $r_{ex}$ represents the multiple of the VBL thicknesses at the inner and the outer boundaries, $\delta_v^i$ and $\delta_v^o$, respectively, that we exclude from the volume-average to ensure the representation of bulk dynamics. There are two common methods of estimating the thickness of the VBL from the radial profile of horizontal velocity (*L20*). One is the "local maxima method" (*L20*), where the distance of the nearest maxima in this profile from the respective boundaries is defined as the VBL thickness, whereas the distance of the intersection of the tangent to the profile at the wall and the tangent at the nearest maxima near the wall is used as VBL thickness in the "linear intersection method" (Gastine *et al.* 2016). In our study, VBL thickness is estimated using the linear intersection method, which has been reported as a better estimate than the local maxima method (Gastine *et al.* 2015). The value of $r_{ex}$ is determined in section 3.1.

We evaluate the scale-dependence of the forces as a function of spherical harmonic degree following Aubert *et al.* (2017). We remove the dynamically irrelevant spherical harmonic degree $l = 0$ from $I^{ff}$ in equation 2.8 (Nicoski *et al.* 2024) and from all the mean force terms in equation 2.7 that represents the hydrostatic balance (Calkins *et al.* 2021). The volume-averaged r.m.s. forces are averaged in time over at least a hundred advective time units after the simulations reach a statistically stationary state (Mound & Davies 2017; Long *et al.* 2020). The calculation of forces follows the methodology reported by Calkins *et al.* (2021) and Nicoski *et al.* (2024), and a couple of cases from Nicoski *et al.* (2024) have been reproduced to validate our force calculation.

## 3. Results

In this section, we investigate the force balances in the thermally-driven RC simulations at $E = 10^{-5}$ reported in Mound & Davies (2017) and *L20*. We start, in section 3.1, with a systematic study to assess the thickness of the region near the boundaries that should be excluded to ensure the representation of bulk dynamics. We explore the mean and fluctuating parts of each component of forces in section 3.2 and their curl in section 3.3. The scale dependence of the forces is investigated in section 3.4 whereas the role of inertia in the force balance in various regimes of RC is considered in section 4. The thermal forcings at which transitions occur from WN to RR to WR behaviours have been previously predicted based on theoretical scaling laws (*L20*) and are marked by vertical lines in the volume-averaged balances presented in the next section. We use representative cases from each of these regimes at $\widetilde{Ra} = 90$ (WN), $\widetilde{Ra} = 1200$ (RR), and $\widetilde{Ra} = 13000$ (WR) to establish the required boundary layer exclusion in section 3.1 and to demonstrate the scale-dependence of the balances in section 3.4. The force balances at $E = 10^{-4}$ and $E = 10^{-6}$ are presented in supplementary figures S1 and S2, respectively, for comparison.

### 3.1. *Boundary layer exclusion*

We begin by assessing the variation of the fluctuating forces (figure 1c,e,g) and their curls (figure 1d,f,h) as a function of $r_{ex}$, which represents the number of VBLs excluded near each boundary from the volume-average. Mean quantities exhibit similar behaviour, and so we focus on fluctuating quantities in this section. A common practice (e.g. Yadav *et al.* 2016; Aubert *et al.* 2017) is to exclude a single VBL before calculating the volume-average (figure 1a,b), which corresponds to $r_{ex} = 1$. We demonstrate the impact of changing $r_{ex}$ in our representative cases from the WN(figure 1c,d), RR (figure 1e,f), and WR (figure 1g,h) regimes. Since viscous forces are strongest near the no-slip boundaries in our simulations, that force term changes most significantly as boundary layers are excluded (figure 1c,e,g). The viscous force term decreases with increasing $r_{ex}$ and converges to its bulk values for $r_{ex} \geqslant 5$ for all cases. The curled viscous force requires more exclusion than the uncurled force to converge to its bulk values (e.g., compare figure 1c and figure 1d), probably due to the presence of an extra spatial gradient that inflates sharp changes near the boundaries.

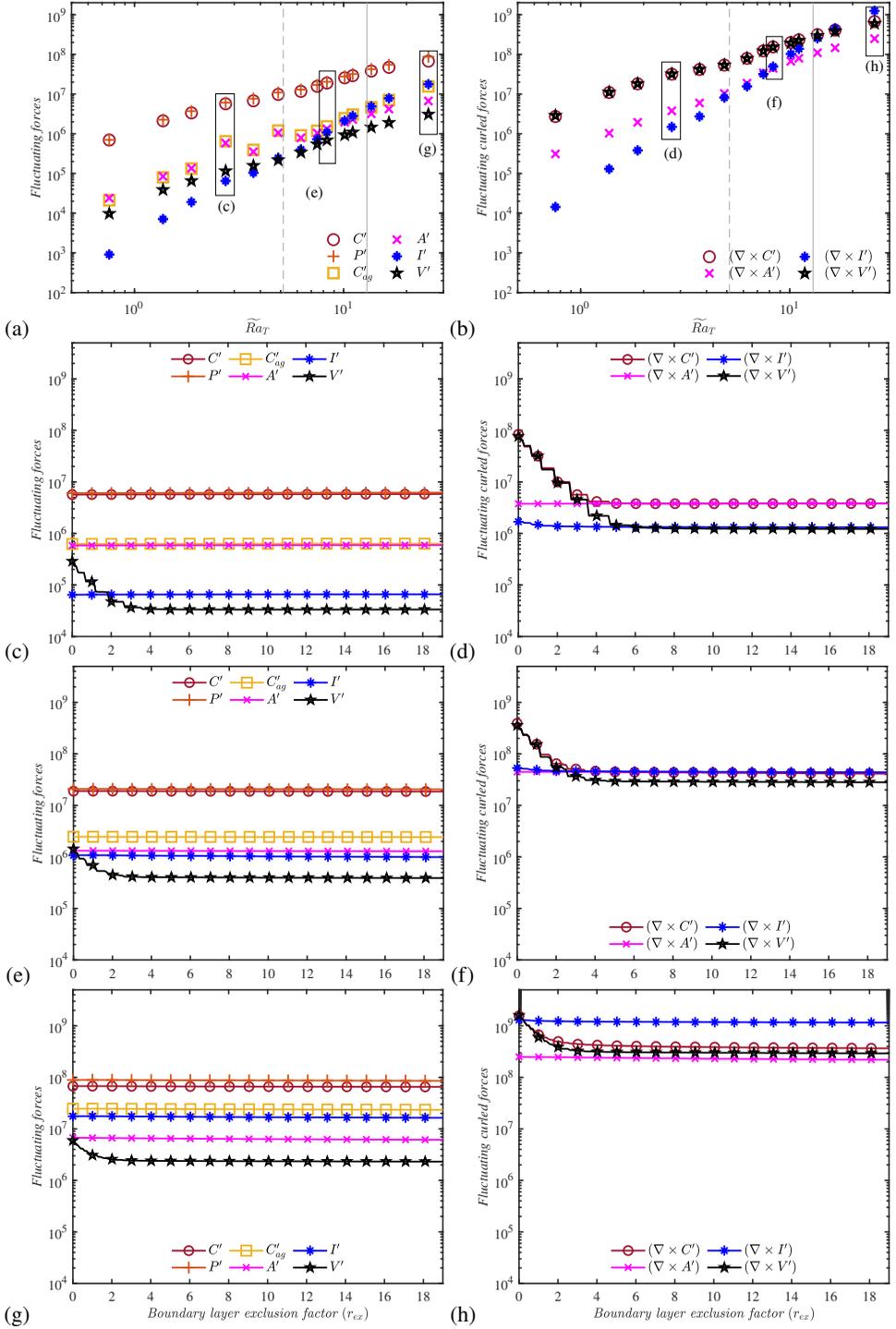

Figure 1: Variation of volume-averaged rms fluctuating forces (left column) and their curls (right column) at $E = 10^{-5}$ with thermal forcing $\widetilde{Ra}_T$ where only single boundary layer thicknesses are excluded ($r_{ex} = 1$) in (a,b), and with boundary layer exclusion factor ($r_{ex}$) in (c-h) for the annotated cases in (a,b). The representative cases from WN, RR, and WR regimes correspond to $\widetilde{Ra} = 150, 1200,$ and $13000$ for $E = 10^{-5}$.

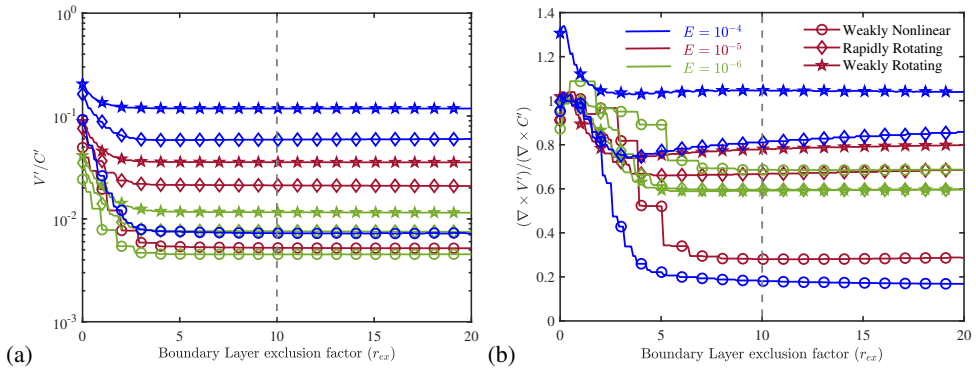

Figure 2: Variation of fluctuating viscous to Coriolis (a) force and (b) curled force ratios with the thickness of the excluded layers as a multiple ($r_{ex}$) of VBL thickness. The representative cases from WN (circles), RR (diamonds), and WR (stars) regimes for three Ekman numbers correspond to $\widetilde{Ra} = 30, 900,$ and $13000$ for $E = 10^{-4}$ (blue), $\widetilde{Ra} = 90, 1200,$ and $13000$ (red) for $E = 10^{-5}$, and $\widetilde{Ra} = 150, 2000,$ and $18000$ for $E = 10^{-6}$ (green).

Crucially, excluding only a single VBL to calculate the forces is generally insufficient to properly capture bulk dynamics in our simulations. In the WN regime, the viscous force is larger than inertia for $r_{ex} = 1$ (figure 1c), while it falls below inertia for $r_{ex} > 2$. Overestimation of the viscous term in the bulk with $r_{ex} = 1$ is particularly problematic when considering curled forces, where it can result in a more than an order of magnitude overestimate (e.g., figure 1d). In the curled forces, a VC balance is obtained for the WN cases using $r_{ex} = 1$, whereas an AC (or TW) balance can be observed for $r_{ex} = 10$ (figure 1d). In the RR regime (figure 1f), the curled force balance changes from VC to IVAC with increasing exclusion. For the WR case (figure 1h), the IVC balance obtained for $r_{ex} = 1$ becomes an IVAC balance with increasing $r_{ex}$. Figure 1 also shows that the ordering of terms in the force and curled balances can differ significantly at low $r_{ex}$, but show better agreement for large $r_{ex}$.

To test the appropriate value of $r_{ex}$ across the entire suite of runs, we plot the ratio of volume-averaged fluctuating viscous to Coriolis terms as a function of $r_{ex}$ in figure 2. This ratio is generally largest for low $r_{ex}$ and decreases with increasing $r_{ex}$. The value of $r_{ex}$ required to obtain converged results (i.e., that do not change upon a further increase in $r_{ex}$) varies somewhat with the parameters but is always larger for the curled terms than the forces. Across our suite of simulations, a converged bulk force balance is always obtained with $r_{ex} \geqslant 5$. The curled force ratio (figure 2b) exhibits slower convergence to the bulk values than the uncurled forces and can require $r_{ex} \geqslant 10$. Following the analysis in figure 2, we exclude ten VBLs (i.e., $r_{ex} = 10$) in all volume averages. This corresponds to $2 - 4$ VBLs if the local maxima method is used instead of a linear intersection method, as the former leads to thicker estimates of the VBL (Gastine *et al.* 2015). For the thickest boundary layers at $\widetilde{Ra} = 30$ and $E = 10^{-4}$, about $34\% - 38\%$ of the gap is excluded from the radial extent of the domain at $r_{ex} = 10$. For the majority of simulations, the total excluded region amounts to less than 20% of the gap width.

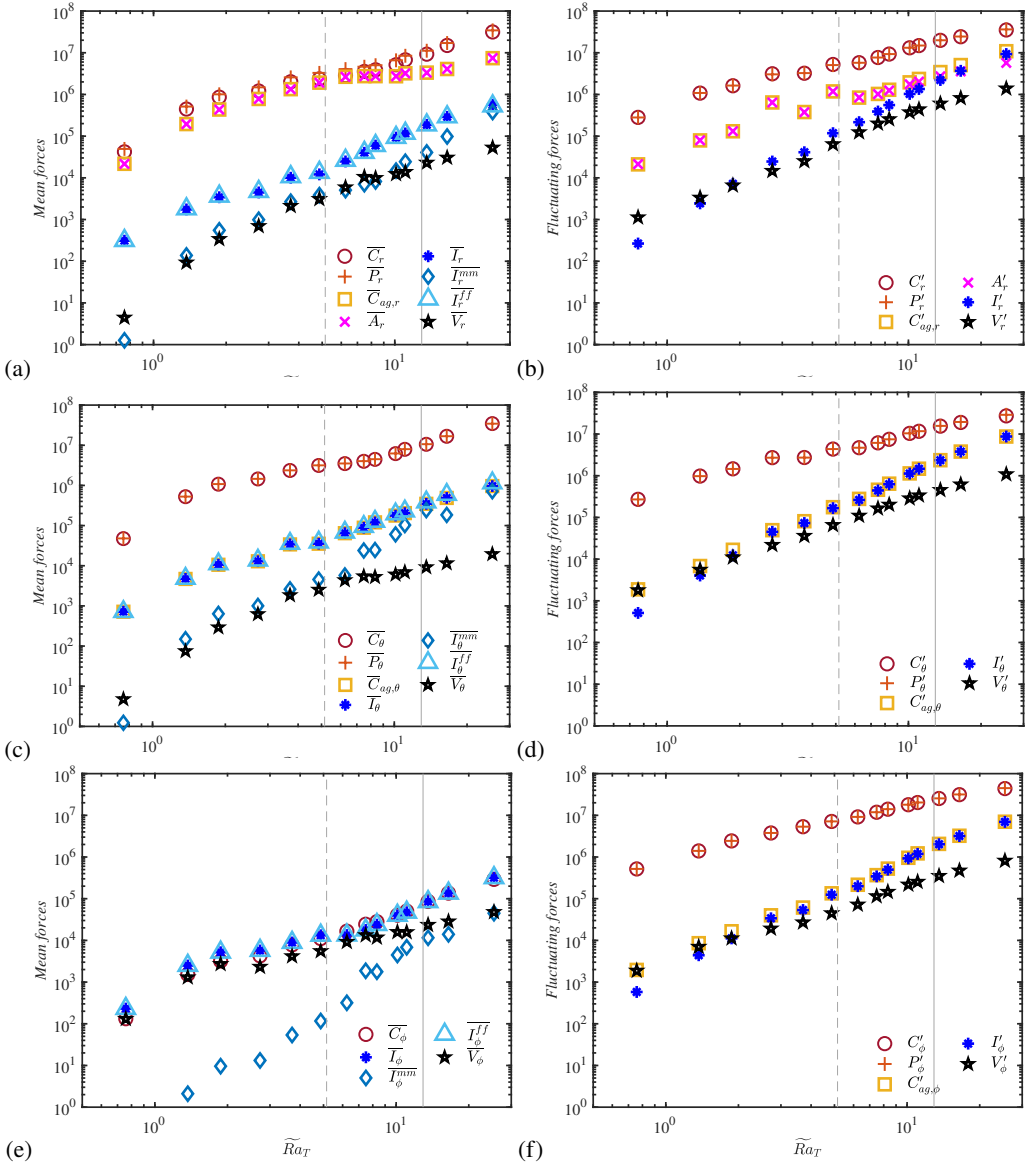

Figure 3: Volume-averaged r.m.s. mean (left column) and fluctuating (right column) force components in $\hat{r}$ (a,b), $\hat{\theta}$ (c,d), and $\hat{\phi}$ (e,f) for $E = 10^{-5}$.

### 3.2. Force balance

The volume-averaged mean and fluctuating forces as a function of thermal forcing, $\widetilde{Ra}_T$, are shown in figure 3a,c,e and 3b,d,f respectively. Because buoyancy acts only radially and the Coriolis force is vertical, there will be a directional anisotropy in the force balances and therefore we consider the radial ($\hat{r}$, figure 3a,b), co-latitudinal ($\hat{\theta}$, figure 3c,d), and azimuthal ($\hat{\phi}$, figure 3e,f) force components separately.

In the radial direction (figure 3a), we find a primary thermal wind (TW) balance between the mean Coriolis, pressure and buoyancy forces, with the buoyancy force gradually becoming subdominant at the highest values of $\widetilde{Ra}_T$ considered. In the $\theta$ direction (figure 3c), the primary balance in the mean forces is quasi-geostrophic (QG, i.e., between Coriolis and pressure forces) for all $\widetilde{Ra_T}$. These mean balances in $\hat{r}$ and $\hat{\theta}$ are the same as the balances reported in recent dynamo simulations (Calkins *et al.* 2021). The residuals of the primary TW balance in $\hat{r}$ and the primary QG balance in $\hat{\theta}$ are balanced by the Reynolds stress ($\overline{I^{ff}}$), which dominates the two components of the total mean advection term ($\bar{I} = \overline{I^{mm}} + \overline{I^{ff}}$) until the largest forcings considered. In the mean $\hat{r}$ and $\hat{\theta}$ balances, viscosity is always subdominant as found by Calkins *et al.* (2021). The primary mean forces in the azimuthal direction (figure 3e), have similar magnitude with the secondary forces in the $\hat{r}$ and $\hat{\theta}$-directions. Owing to our choice of averaging, there is no mean pressure gradient or buoyancy force in the azimuthal direction. Therefore, the Coriolis force is balanced by inertia and viscous forces at low $\widetilde{Ra_T}$ (*IVC* balance), whereas an *IC* balance dominates at high $\widetilde{Ra_T}$.

In the fluctuating forces in the right column (figures 3b,d,f), the QG balance between Coriolis and pressure forces dominates in all directions. At low $\widetilde{Ra_T}$ in the radial direction, there is a secondary $AC_{ag}$ balance between ageostrophic Coriolis force and buoyancy with viscosity approximately an order of magnitude weaker. At high $\widetilde{Ra_T}$ in the radial direction, there is a secondary $IAC_{ag}$ balance with viscosity still subdominant but less than an order of magnitude weaker. In the $\theta$ and $\phi$-directions, the ageostrophic Coriolis force is balanced by both viscous and inertial forces for low $\widetilde{Ra_T}$ ($IVC_{ag}$ balance), with the relative importance of viscosity weakening as $\widetilde{Ra_T}$ increases.

The behaviour of the fluctuating radial forces can be compared with the free-slip and fixed temperature simulations of Nicoski *et al.* (2024) (see figure 16(b) in their paper). For both studies, the primary radial balance is QG, while the secondary balance is $AC_{ag}$ at low thermal forcings and $IAC_{ag}$ with a weakly subdominant viscous contribution at higher thermal forcings. Though the viscous force can be higher than inertia for the lowest $\widetilde{Ra_T}$ values considered here (figure 3b), it is the smallest force in both studies when the same range of $\widetilde{Ra_T}$ is considered. However, the two studies find different dominant contributions to the inertia force. Figure 3b shows that $I'$ is dominated by $I^{ff}$, which balances $C_{ag}$ and $A'_r$ for large $\widetilde{Ra_T}$. In contrast, Nicoski *et al.* (2024) found that the mean-fluctuating inertia ($I^{mf}$) dominates the non-linear advection terms, which, combined with the time derivative $TD$, balances $C_{ag}$ and $A'_r$ for large $\widetilde{Ra_T}$. This difference probably arises from the strong mean zonal flows in the simulations of Nicoski *et al.* (2024), owing to the use of free-slip conditions.

The partitioning of the forces into mean ($m = 0$) and fluctuating components ($m \neq 0$) brings out distinct balances (comparing the left and right panels of figure 3). For example, the radial force balance exhibits a primary mean TW balance and a primary fluctuating QG balance. The unpartitioned (i.e., mean+fluctuating) radial force balance has a primary QG balance that exceeds a subdominant $AC_{ag}$ balance by a factor of $2 - 3$, essentially averaging the large- and small-scale dynamics.

The azimuthal component of azimuthally averaged forces has been considered in previous studies (Sheyko *et al.* 2018; Menu *et al.* 2020) since it removes the pressure gradient, although there is no buoyancy force in this direction. Our analysis indicates that this representation (see figure 3e for an example) does not reflect the balance of mean forces in the $\hat{r}$ and $\hat{\theta}$ and also does not correspond to the balances of the fluctuating part of the forces.

### 3.3. *Curled force balance*

Though the force balance provides useful insights into the dynamics, all forces in equation 2.2 are non-solenoidal and, therefore, will have gradient portions. These gradient portions of the forces are balanced by the pressure gradient term, which plays no role in the dynamics (Hughes & Cattaneo 2019; Teed & Dormy 2023). Our approach to removing these gradients is to take the curl of the momentum equation. We partition the curled forces into mean (figure 4a,c,e) and fluctuating (figure 4b,d,f) parts and separately consider the radial (figure 4a,b), co-latitudinal (figure 4c,d), and azimuthal (figure 4e,f) components. Since the buoyancy force is radial, its curl acts only in the angular directions ($\theta$ and $\phi$), making the curled balance inherently anisotropic.

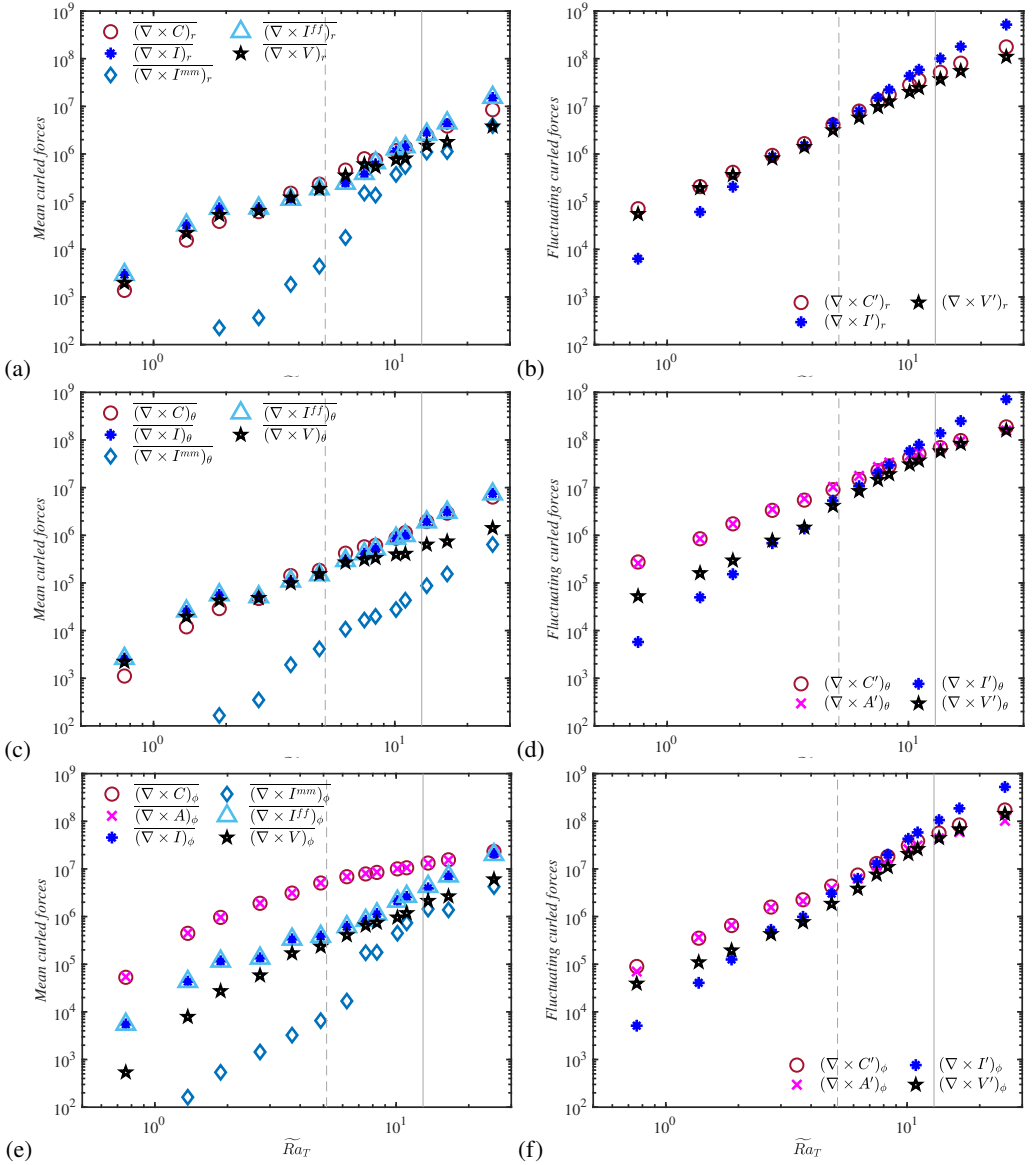

Figure 4: Volume-averaged r.m.s. mean (left column) and fluctuating (right column) curled force components in $\hat{r}$ (a,b), $\hat{\theta}$ (c,d), and $\hat{\phi}$ (e,f) for $E = 10^{-5}$.

In the $\hat{r}$ and $\hat{\theta}$-directions, we find a primary mean balance between the Coriolis, inertial, and viscous terms (*IVC* balance), with the viscous contribution weakening at the highest values of $\widetilde{Ra}_T$ considered. In the $\hat{\phi}$ direction, the mean balance is a thermal wind, except at the highest values of $\widetilde{Ra}_T$ considered where the mean curled inertia also enters this primary balance. A thermal wind arises only in the mean azimuthal component because our choice of averaging causes the $\theta$-component of the mean curled buoyancy force to vanish (since $\partial \overline{T}/\partial \phi = 0$). Aubert (2005) also found a TW balance in the azimuthally averaged curled force balance for non-magnetic simulations at $E = 10^{-4} - 10^{-5}$.

The curled fluctuating forces in $\hat{\theta}$ and $\hat{\phi}$ exhibit a primary balance between Coriolis, buoyancy, and viscous terms ($VAC$ balance) at low $\widetilde{Ra}_T$, while the inertial force gradually enters this balance with increasing $\widetilde{Ra}_T$. A similar trend is observed in the radial balance, except for the omission of the buoyancy term. We note here that the radial curled force balance, as reported by Dormy (2016), may not represent the curled force balance in the other directions.

As with the forces, partitioning curled terms into mean and fluctuating components brings out distinct balances that would be obscured if only the unpartitioned curled terms were considered. In particular, because the fluctuating curled quantities have much higher amplitude than the mean curled quantities, an unpartitioned (i.e., mean+fluctuation) curled force representation would simply show the fluctuating small-scale balance. The balance obtained from the total magnitude of the three mean curled components follows the balance in the $\hat{\phi}$ direction shown in figure 4e, while the balance obtained from the total magnitude of the three fluctuating curled components reflects the balances in the $\hat{\theta}$ and $\hat{\phi}$ directions shown in figures 4d,f.

### 3.4. *Scale-dependent force balance*

In figure 5, we compare the scale dependence of the total magnitude of the fluctuating forces (left column) and their curls (right column) with the corresponding scale-integrated representation (first row). For brevity, we again select one case from each of the WN (c,d), RR (e,f), and WR (g,h) regimes, corresponding to $\widetilde{Ra} = 90, 1200,$ and $13000$, respectively (these cases are highlighted in figure 5a,b). The regime boundaries are shown using dashed and solid vertical lines, following the analysis of $L20$, as described in detail in the next section. Although the mean forces are also dependent on the spherical harmonic degree, we focus on the fluctuating forces because of their more direct correspondence with the convective motions and the heat transfer behaviour.

Based on the scale-integrated fluctuating forces, all simulations have a QG primary balance (figure 5a). This primary force balance holds across all scales in the WN and RR regimes (figures 5c,e), while inertia enters the primary balance at the small scales (i.e., large $l$) in the WR regime (figure 5g). The secondary force balance in the WN and RR regimes is characterised by an $AC_{ag}$ balance between buoyancy and ageostrophic Coriolis at large scales, with inertia entering at small scales (figure 5e). In the WR regime, there is a secondary $IAC_{ag}$ balance with a significant inertial contribution at all scales. The scale dependence of forces in the RR regime (figure 5e) is similar to the force spectra reported in Schwaiger *et al.* (2020) (see figure 7(a) in their paper) and is often referred to as an $QG - IAC_{ag}$ balance.

Previous studies have attempted to relate crossings of scale-dependent forces (such as the crossing of buoyancy and inertia forces at $l \sim 22$ in figure 5e) to dynamically relevant length scales (e.g. Schwaiger *et al.* 2020). However, for high $\widetilde{Ra}_T$ in the WR regime (figure 5g), the observed crossings occur at $l \sim 3$, which is much smaller than the dominant wavenumber of the flow ($l \sim 15$, based on the peak of the kinetic energy spectra). Therefore, it may not always be possible to relate crossings of scale-dependent forces to dynamically significant length scales.

Compared to the scale-dependent forces, the scale-dependent curled balances do not produce a clear separation of balances (i.e., primary/secondary/tertiary). In the WN regime (figure 5d), an *AC* balance is evident at large scales, while viscous and inertia forces enter the balance at small scales. The curled force magnitudes in the RR regime (figure 5f) are comparable at all scales, leading to an *IVAC* (Inertia-Viscous-Archimedean-Coriolis) balance in the scale-integrated representation (figure 5b). In the WR regime (figure 5h), an *IC* balance is observed at large scales, while the inertia dominates the balance at small scales.

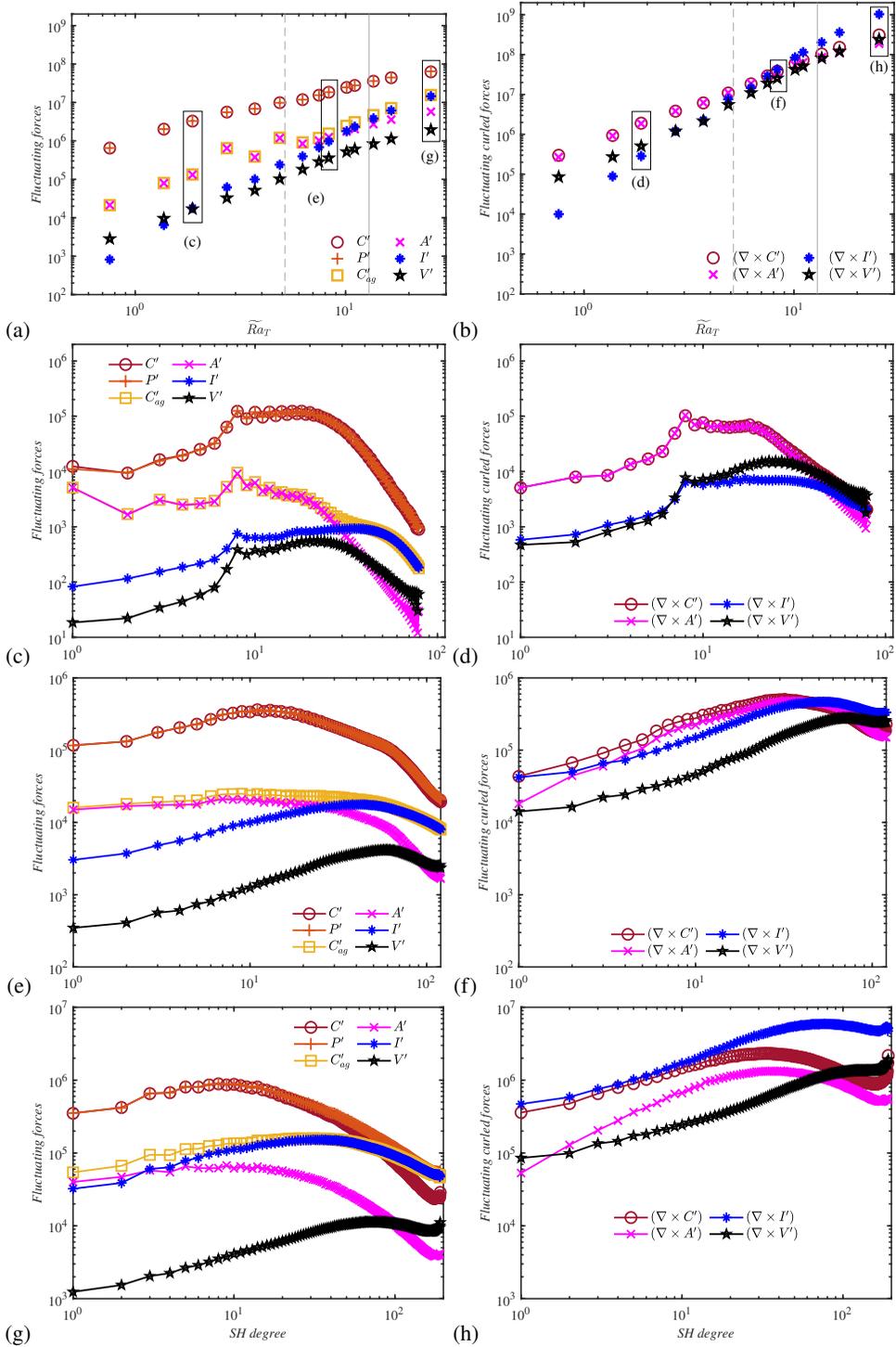

Figure 5: Volume-averaged r.m.s. of the total magnitude of fluctuating forces (left column) and their curls (right column) at $E = 10^{-5}$ as a function of thermal forcing $\widetilde{Ra}_T$ in (a,b) and with SH degree ($l$) in (c-h) for the exemplar cases highlighted in (a,b).

Figure 5 compares scale-integrated and scale-dependent representations of dynamical balances in our simulations. Integrated forces do not capture crossover scales where they exist or the general decrease of the buoyancy force and increase of the inertial force with increasing $l$ (e.g., figure 5e). Furthermore, integrated quantities generally do not reflect the force balance at the smallest scales of the solution. Nevertheless, the integrated representation quantitatively captures the overall ordering of forces in a single measure that can be easily compared across a large suite of simulations. In comparison, the curled balances are comparatively less scale-dependent, neither exhibiting a clear ordering of the forces nor any distinct cross-over scales (see also Teed & Dormy 2023). Therefore, a scale-integrated analysis (figure 5b) is sufficient to represent the curled forces.

## 4. Summary of Dynamical Balances and Comparison to Regimes of *L*20

Table 1 presents a qualitative summary of the different balances that we found within our suite of simulations depending on whether we considered forces or curled forces, total magnitudes or individual vector components, or partitioning into azimuthally mean and fluctuating contributions. We reiterate here that our analysis reflects bulk dynamics, with the volume averages obtained after removing ten VBLs from each boundary of the domain. Balances similar to those described in detail above for $E = 10^{-5}$ are also found at $E = 10^{-4}$ and $E = 10^{-6}$; the main difference is that the separation between the primary and secondary balances, denoted by a dash (−) in the table, increases with decreasing Ekman number (e.g., compare figure 5 with figures 3(a) in both S1 and S2). We have tried to denote "balances" which are groupings of two or more terms that are separated by an order of magnitude in amplitude from other terms; however, such a large separation is not always present. Changes in the balances with $\widetilde{Ra}_T$ are indicated by a right arrow (→), which indicates the general trends with increased thermal forcing but not the specific values at which the balances change. Therefore, table 1 is only a general description of the complex variations in force balances amongst our suite of simulations. In general, increased thermal forcing results in an increase in the relative importance of inertial terms in the balances.

We now compare transitions in the force and curled force balances to previous predictions of regime transitions based on scaling laws (*L*20). We put emphasis on the $E \leqslant 10^{-5}$ cases as they are more appropriate for comparing with the asymptotic scaling theories used by *L*20. They defined the WN-RR regime transition (dashed vertical lines in figures 3–5) as $Ra = 8Ra_c$ based on the observed gradual departure from the linear $Nu - 1 \propto Ra/Ra_c - 1$ scaling expected just above onset. In the WN regime, they found that the simulated flow lengthscale $\ell$ and convective Reynolds number $Re_c$ follow the predictions of VAC theory. *L*20 defined the RR-WR transition (solid vertical lines in figures 3–5) based on the condition $RaE^{8/5} \sim O(1)$ of Julien *et al.* (2012a) above which the thermal boundary layers lose geostrophic balance. They found scalings for $\ell$ and $Re_c$ close to but statistically different from the predictions of IAC theory.

Figures 3 and 4 show that mean forces exhibit no changes in primary or secondary balances over the range of $\widetilde{Ra}_T$ considered and hence do not conform to the regime transitions found by *L*20. This is expected since *L*20 defined transitions based on quantities that depend strongly on convective fluctuations such as $Nu$, $\ell$, and $Re_c$. In the fluctuating forces, the WN-RR transition correlates with viscous and inertial terms coming into approximate balance with the ageostrophic and buoyancy terms that comprise the secondary balance (figure 5a), while in the fluctuating curled forces this transition arises when the magnitude of the nonlinear advection term becomes comparable to the Coriolis, buoyancy, and viscous terms (figure 5b). In the fluctuating forces, the RR-WR transition broadly correlates with the amplitude of the viscous term falling below that of the secondary balance, while in the fluctuating curls, this transition appears to correlate with the amplitude of the inertial term rising above the remaining terms. However, the total fluctuating forces and curled forces do not suggest an exact value of $\widetilde{Ra}_T$ where regime transitions occur. Indeed, these quantities exhibit gradual changes with $\widetilde{Ra}_T$ and $E$ (e.g. figures 5a,b and figures 3a,b in both S1 and S2) and hence any transition inferred from them is necessarily broad rather than abrupt.

| | Forces | | Curled forces | |
|---|---|---|---|---|
| component | Mean | Fluctuating | Mean | Fluctuating |
| r | $TW - I(TW)_{res}$ | $QG - (AC_{ag} \to IAC_{ag})$ | $IVC$ | $VC \to IVC$ |
| $\theta$ | $QG - IC_{ag}$ | $QG - (IVC_{ag} \to IC_{ag})$ | $IVC$ | $VAC \to IVAC$ |
| $\phi$ | $IVC \to IC$ | $QG - (IVC_{ag} \to IC_{ag})$ | $TW - IV(TW)_{res} \to IAC$ | $VAC \to IVAC$ |
| tot | $TW - I(TW)_{res}$ | $QG - (AC_{ag} \to IAC_{ag})$ | $TW - IV(TW)_{res} \to IVAC$ | $VAC \to IVAC$ |

Table 1: Summary of force and curled force balances in our simulations. The $ACP$ balance of forces (or the $AC$ balance of curled forces) is referred to as a thermal wind (TW) balance, while the residual of these forces is designated here as $(TW)_{res}$. Similarly, the primary balance between Coriolis and pressure gradient forces is denoted as a $QG$ balance. Primary and secondary force balances are separated by a dash (−), while the changes in the balance with increasing thermal forcing ($\widetilde{Ra}_T$) are designated with a right arrow (→).

Table 2 summarises the general character of the balances in the regimes defined by $L20$, calculated using the total magnitude of the mean (supplementary material S3) and fluctuating forces and curled forces. $L20$ inferred a $VAC$ balance in the WN regime and an $IAC$ balance in the RR regime using scaling theory based on the curled force balance. In the WN regime, the calculated fluctuating curled force balance is $VAC$ at low $\widetilde{Ra}_T$, transitioning to an $IVAC$ balance as $\widetilde{Ra}_T$ increases (figure 5b). This behaviour is broadly consistent with the assumptions of $L20$. In the section of the RR regime accessed by our simulations, the calculated fluctuating curled force balance is $IVAC$ rather than the IAC balance assumed by $L20$. The viscous force is also significant in the force balance (figure 5a), though it remains smaller than the other forces in the RR regime. Similar behaviour of the viscous force can also be observed at $E = 10^{-4}$ and $E = 10^{-6}$ (see figures 3a,b in both S1 and S2). Reconciling the calculated dynamical balances in the RR regime with other flow diagnostics such as $\ell$, $Nu$, and $Re$ must await a scaling theory for the $IVAC$ regime.

In summary, the fluctuating force and curled force balances exhibit smooth variations over the range of $\widetilde{Ra}_T$ and $E$ considered, reflecting gradual rather than abrupt changes in the dynamics. Broadly speaking, it appears that the RR regime as defined by $L20$ corresponds to a range of $\widetilde{Ra}_T$ where inertial effects enter the primary fluctuating curled force balance (or the secondary fluctuating force balance). This observation motivated us to seek a single parameter to characterise the changing dynamics with $\widetilde{Ra}_T$. However, it is difficult to find a single quantity that adequately represents the transitions in heat transport and flow behaviour identified by $L20$. This is perhaps unsurprising given that even the simple $VAC$ scaling laws used by $L20$ are defined by at least two parameters, while the dominate dynamical balances identified in table 2 involve at least three terms. We, therefore, classify these balances by introducing two parameters based on our calculated dynamical balances.

| | Forces | | Curled forces | |
|---|---|---|---|---|
| Regime | Mean | Fluctuating | Mean | Fluctuating |
| WN | $TW - I(TW)_{res}$ | $QG - AC_{ag}$ | $TW - IV(TW)_{res}$ | $VAC \to IVAC$ |
| RR | $TW - I(TW)_{res}$ | $QG - IAC_{ag}$ | $TW - IV(TW)_{res} \to IVAC$ | $IVAC$ |
| WR | $TW - I(TW)_{res}$ | $QG - IAC_{ag}$ | $IVAC$ | $IVAC$ |

Table 2: Summary of force and curled force balances in the regimes of RC simulations as predicted by $L20$. The balances in total force magnitudes (equation 2.9) have been used here. The abbreviations used here are the same as described in table 1. Primary and secondary force balances are separated by a dash ($-$), while the changes in the balance with increasing thermal forcing ($\widetilde{Ra}_T$) are designated with a right arrow ($\to$).

Figure 6 shows a quantitative comparison of our dynamical balances with the regime diagram of $L20$ (their figure 14). To classify the balance in the $E - Ra/Ra_c$ parameter space, we are motivated by the observation that inertia varies most strongly with $\widetilde{Ra}_T$ in our simulations (figure 5). We therefore introduce two new measures based on the total magnitude of the fluctuating forces and the curled forces (figure 5a,b) that assess the role of inertia in the force balance.

To measure the degree of geostrophy in the balance we define a force ratio,

$$\mathcal{F}_{I/C} = \frac{I'}{C'}. \qquad (4.1)$$

We find that the value of $\mathcal{F}_{I/C} = 0.1$ can be used to demarcate the simulations belonging to WR regime as demonstrated in figure 6. Here, $\mathcal{F}_{I/C} > 0.1$ (open symbols) can describe almost all simulations that fall in the WR regime, as compared to the simulations in the range $\mathcal{F}_{I/C} \leqslant 0.1$ (filled symbols) that mostly fall inside the WN and RR regimes.

To measure the role of inertia in the curled balance we further define a curled force ratio,

$$C\mathcal{F}_{I/C} = \frac{\nabla \times I'}{\nabla \times C'}. \qquad (4.2)$$

The symbols in figure 6 are coloured by $log_{10}(C\mathcal{F}_{I/C})$. We emphasize again that the role of all terms, including inertia, in the dynamical balances change gradually, and hence values of $\mathcal{F}_{I/C}$ and $C\mathcal{F}_{I/C}$ that reflect regime transitions must be chosen arbitrarily. Nevertheless, the RR regime can be characterized by the combination of $\mathcal{F}_{I/C} \lesssim 0.1$ and $log_{10}(C\mathcal{F}_{I/C}) \approx 0$ (i.e., $\mathcal{CF}_{I/C} \approx 1$), which reflects the primary QG balance in the fluctuating forces and the IVAC balance in the fluctuating curled forces, respectively (table 2).

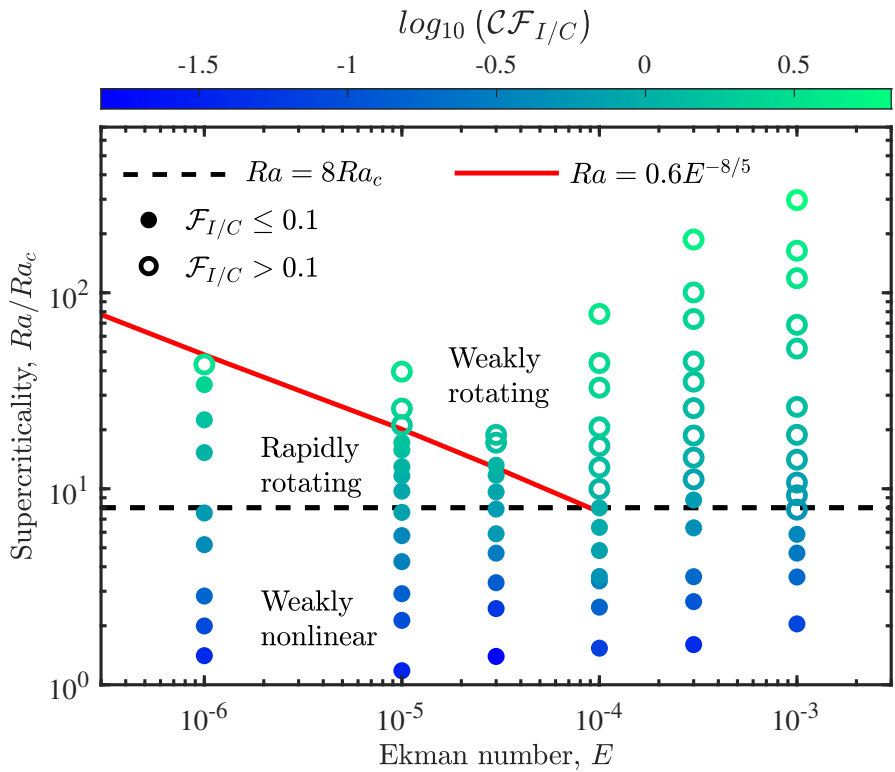

Figure 6: Regime diagram with predicted regime boundaries from the analysis of *L*20. The primary force balance is classified as QG when $\mathcal{F}_{I/C} \leqslant 0.1$ (filled symbols), and non-QG when $\mathcal{F}_{I/C} > 0.1$ (open symbols). The markers are coloured by $log_{10}(\mathcal{CF}_{I/C})$. *L*20 found the WN-RR regime transition corresponded to the condition $Ra/Ra_c = 8$, and the RR-WR transition corresponded to the condition $Ra = 0.6E^{-8/5}$ of Julien *et al.* (2012*b*).

## 5. Discussion and Conclusions

We have analysed different representations of dynamical balances in simulations of spherical shell rotating convection. The radial, co-latitudinal and azimuthal components of the forces have been considered separately to demonstrate the anisotropic nature of dynamical balances. We also partition the forces into azimuthally averaged mean and corresponding fluctuating parts that exhibit distinct balances. Furthermore, the curled force components are also analyzed to investigate the solenoidal force balance. The utility of a scale-dependent representation of curled and uncurled forces has been addressed. Our main findings are presented in figures 3–5 and table 1 and can be summarised as follows:

(i) The bulk curled force balance depends critically on the number of VBLs that are removed near the upper and lower boundaries. We find that removing ten VBLs from each boundary provides a robust estimate of the curled force balance that is broadly consistent with the balance obtained from calculating forces.

(ii) Mean and fluctuating forces and curled forces exhibit distinct balances, consistent with the results of Calkins *et al.* (2021) and Nicoski *et al.* (2024). In particular, the primary mean force and curled force balances are TW, while the primary fluctuating force (curled force) balance is $QG$ ($IVAC$).

(iii) Radial, co-latitudinal, and azimuthal forces exhibit distinct balances as found by Calkins *et al.* (2021) and Aubert (2005) for dynamo simulations. For example, mean forces exhibit a primary thermal wind ($TW$) balance in the radial direction, quasi-geostrophic ($QG$) balance in the latitudinal direction, and inertial-viscous-Coriolis balance in the longitudinal direction. A total force magnitude representation underestimates the role of buoyancy compared to the radial balance.

(iv) In the scale-dependent balances, the separation of magnitude between the forces decreases when a curl operation is performed. Cross-over scales are observed in some but not all force balances and are not observed in curled force balances, consistent with the results of Teed & Dormy (2023). The curled forces are only weakly scale-dependent and, therefore, suitably represented by scale-integrated quantities.

(v) Transitions in fluctuating force and curled force balances are broadly consistent with the three regimes of RC obtained by $L20$. However, the relative importance of forces (and their curls) varies gradually with thermal forcing rather than exhibiting any abrupt changes and, therefore, does not define precise values of transition parameters.

(vi) The rapidly rotating (RR) regime broadly corresponds to a range of thermal forcing where inertia is of comparable magnitude to the other terms in the primary curled force balance. We find an $IVAC$ rather than an $IAC$ balance in the fluctuating curled forces in the RR regime for the investigated parameter regime (see table 1). Also, the viscous force increases with increasing thermal forcing, consistent with the results of Nicoski *et al.* (2024).

The dynamical balances in table 1 can be compared to results obtained in previous studies. The mean force balance in the dynamo simulations of Calkins *et al.* (2021) with no-slip boundary conditions is TW in the $\hat{r}$ direction and QG in the $\hat{\theta}$ direction, consistent with our results. The mean curled forces in a dynamo simulation with no-slip conditions (Aubert 2005) is also in a TW balance in the azimuthal direction as in our non-magnetic simulations (4a). This indicates that the primary mean balance in $\hat{r}$ and $\hat{\theta}$ is consistent between non-magnetic and dynamo simulations and, hence, is relatively unaffected by the presence of a magnetic field.

The radial fluctuating force behaviour in our no-slip simulations is consistent with the results of Nicoski *et al.* (2024), who observed a primary QG balance in simulations with stress-free boundary conditions. This indicates that the fluctuating balance in non-magnetic RC is not sensitive to the velocity boundary conditions. Notably, the fluctuating viscous force remains non-negligible within our suite of simulations, similar to the findings of Nicoski *et al.* (2024). Indeed, in the fluctuating curled forces (figure 5b), the viscous term is always part of the dominant balance, even at $E = 10^{-6}$ (see fig 3b in Supplementary Material S2).

Our scale-dependent force balance in the RR regime is consistent with the balance reported in a previous non-magnetic simulation (Schwaiger *et al.* 2020). Although the scale-dependent force balance does not always exhibit a clear cross-over between forces in all regimes in RC, we find a cross-over between buoyancy and inertia forces in the secondary balance in the RR regime (figure 5e and figures 3e in S1 and S2). However, as observed previously by Teed & Dormy (2023), such cross-overs are not found in the scale-dependent curled balance (figure 5f and figures 3f in S1 and S2). This can be attributed to the separations among the terms, which reduce owing to the curl operation that removes the dynamically irrelevant gradient part. Also, the viscous force, which is important only in the small-scale force balance (figure 5e), is significant at all scales in the curled balance (figure 5f). This behaviour of the viscous force remains the same for a lower Ekman number ($E = 10^{-6}$, see Figure 3f in S2). Whether the asymptotic separation among various forces, as demonstrated in the numerical simulations of Nicoski *et al.* (2024), can also be demonstrated for curled forces, requires future simulations at lower Ekman numbers ($E < 10^{-6}$).

Our analysis does not take into account the temporal variation of dynamical balances (see e.g. Schaeffer *et al.* 2017), though previous studies suggest that these variations are small (Aubert *et al.* 2017). We also do not consider dynamical balances in the boundary layers since we aim to characterise the bulk dynamics that would ultimately be used to extrapolate to the conditions of planetary interiors and can be compared to available observations. Nevertheless, the calculated bulk dynamical balances form a basis for comparison with different theoretical analyses of rotating convection (e.g. Aubert *et al.* 2001; Calkins *et al.* 2021; Nicoski *et al.* 2024). Similar force calculations can be useful to study various dynamical regimes of convection with $Pr \neq 1$ (Guzmán *et al.* 2021; Calkins *et al.* 2012), double-diffusive convection (Tassin *et al.* 2021), or geodynamo simulations (Calkins *et al.* 2021; Mound & Davies 2023).

**Acknowledgements.** SN, CJD and JEM are supported by Natural Environment Research Council research grant NE/W005247/1. ATC and CJD are supported by NE/V010867/1. This work used the ARCHER UK National Supercomputing Service (http://www.archer.ac.uk) and ARC2, part of the High Performance Computing facilities at the University of Leeds, UK. We would like to thank J.A. Nicoski and M.A. Calkins for providing the data to validate our force calculations.

**Declaration of interests.** The authors report no conflict of interest.

**Author ORCID.**
S. Naskar, https://orcid.org/0000-0003-0445-8417;
C. J. Davies https://orcid.org/0000-0002-1074-3815;
J. E. Mound https://orcid.org/0000-0002-1243-6915;
A.T. Clarke, https://orcid.org/0000-0003-2128-0016

## Appendix A. Table of results

A summary of the characteristics of the three new simulations performed in this study is reported in table 3. The model resolution, input parameters and selected output diagnostic quantities complement table 5 of Appendix B of Mound & Davies (2017) and table 12 in the appendix of $L$20. Here $N$ is the numerical resolution, which equals both the number of radial points and the maximum spherical harmonic degree and order. $N_{\delta i}$ and $N_{\delta o}$ are the number of radial points within the VBLs at the inner and outer boundary, respectively, where the VBL thicknesses are estimated from the linear intersection method as described in section 2.3. Definitions of the Ekman and modified Rayleigh numbers are given in section 2.1. The Reynolds number is defined as $Re = U^*h/\nu = U = \sqrt{2KE/V_s}$, where $U$ is the non-dimensional velocity (the asterisk indicates dimensional quantity), $V_s$ is the shell volume, and $KE = \iiint_{V_s} \boldsymbol{u}.\boldsymbol{u}\,dV$ is the kinetic energy integral. $Re_{pol}$ is found by retaining only the poloidal velocity in the kinetic energy integral. $Re_{zon}$ is found by retaining only the terms with spherical harmonic order $m = 0$, from the spherical harmonic expansion of the toroidal velocity in the kinetic energy integral. $\mathcal{P}$ is the time average of the buoyancy production throughout the shell, and $\epsilon_U$ is the time average of the viscous dissipation throughout the shell.

| $\widetilde{Ra}$ | $Nu$ | $Re$ | $Re_{pol}$ | $Re_{zon}$ | $\mathcal{P}$ | $\epsilon_U$ | $N$ | $N_{\delta i}$ | $N_{\delta o}$ |
|---|---|---|---|---|---|---|---|---|---|
| 350 | 1.65 | 165.1 | 96.4 | 43.0 | 1.67138e+09 | 1.67131e+09 | 144 | 9 | 7 |
| 550 | 1.78 | 229.9 | 113.9 | 62.2 | 2.80715e+09 | 2.80718e+09 | 160 | 10 | 7 |
| 30000 | 17.06 | 2437.7 | 1211.0 | 1186.6 | 1.76636e+10 | 1.22597e+10 | 384 | 19 | 18 |

Table 3: Summary of the three new runs at $E = 10^{-6}$

# Supplementary Material S1: Forces balance at $E = 10^{-4}$

October 4, 2024

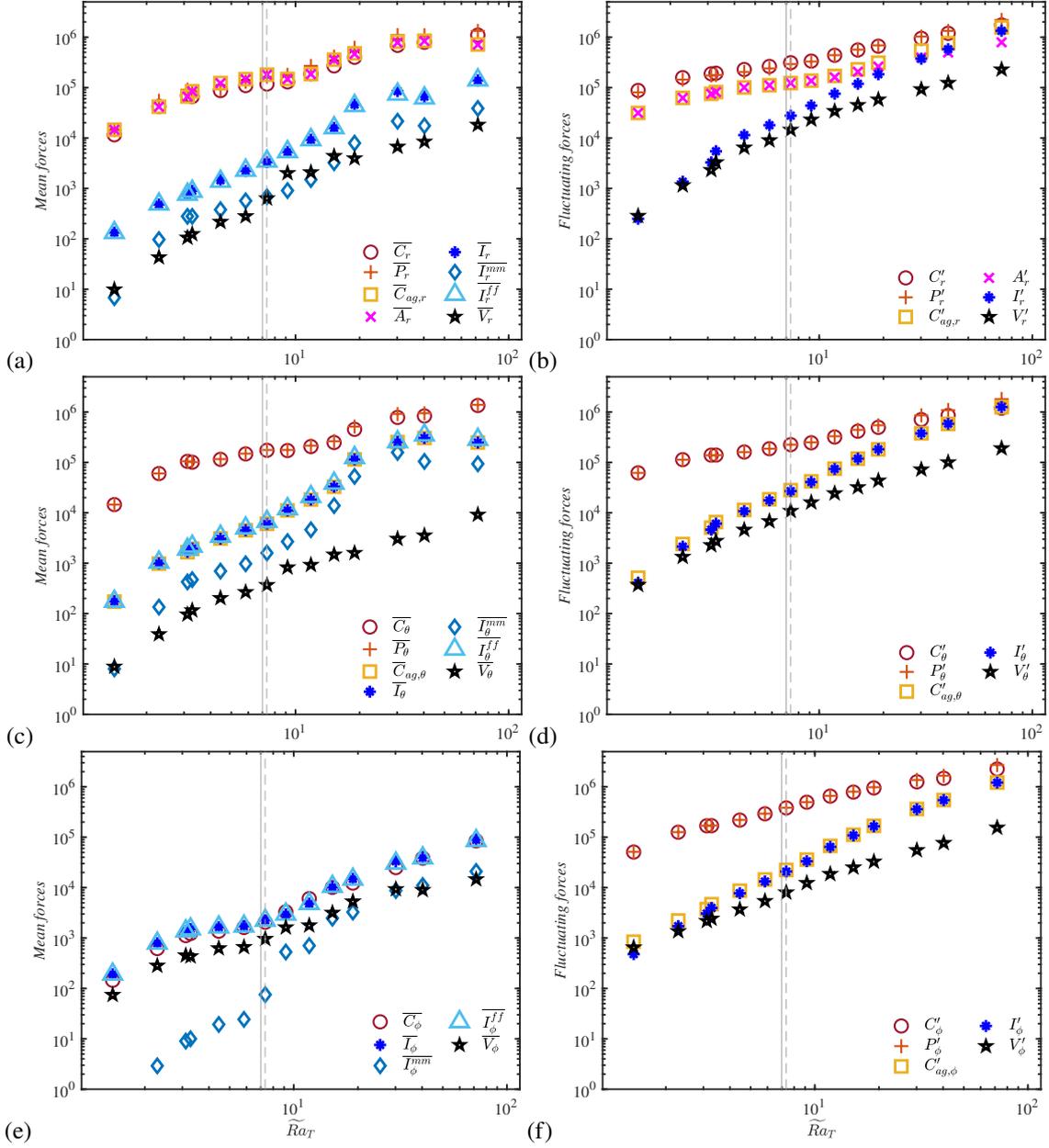

Figure 1: Volume-averaged r.m.s. mean (left column) and fluctuating (right column) force components in $\hat{r}$ (a,b), $\hat{\theta}$ (c,d), and $\hat{\phi}$ (e,f) for $E = 10^{-4}$. The dashed vertical line represents the thermal forcing where a transition from WN to RR regime happens according to the scaling predictions of Long *et al.* (2020).

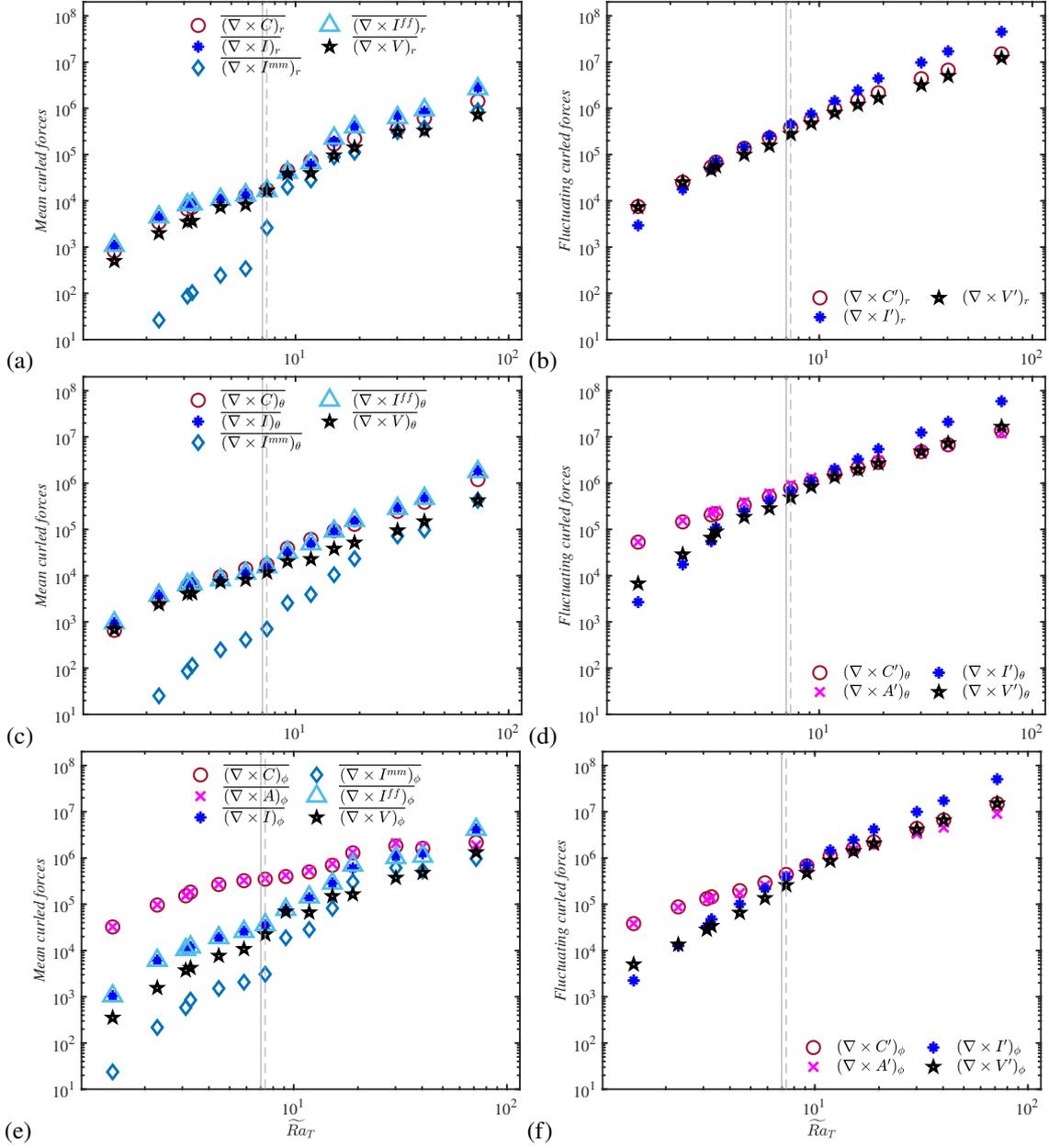

Figure 2: Volume-averaged r.m.s. mean (left column) and fluctuating (right column) curled force components in $\hat{r}$ (a,b), $\hat{\theta}$ (c,d), and $\hat{\phi}$ (e,f) for $E = 10^{-4}$. The dashed (solid) vertical lines represent the thermal forcing where a transition from WN to RR (RR to WR) regime happens according to the scaling predictions of Long *et al.* (2020).

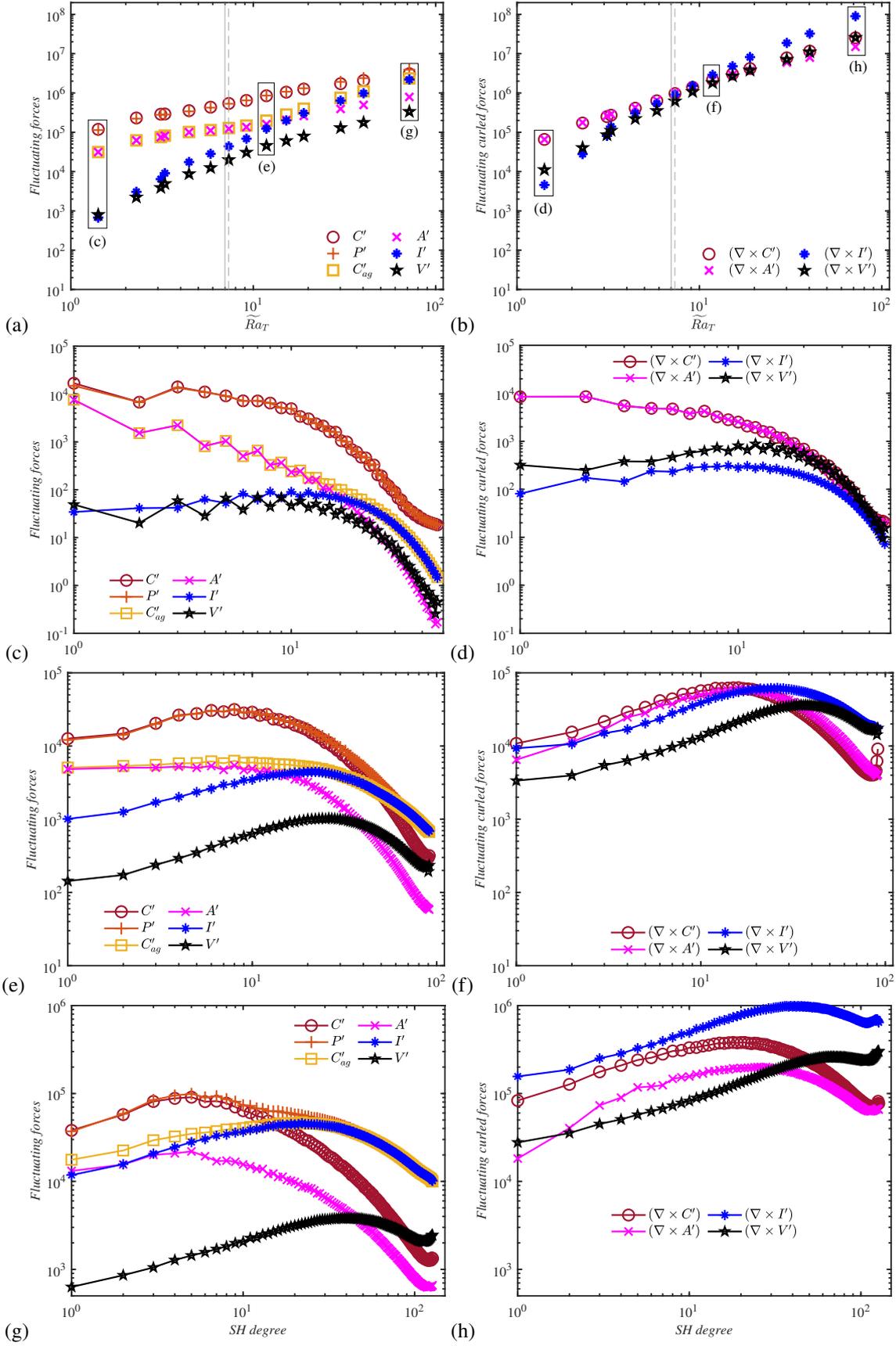

Figure 3: Volume-averaged r.m.s. fluctuating forces (left column) and their curls (right column) at $E = 10^{-4}$ as a function of thermal forcing $\widetilde{Ra}_T$ in (a,b) and spherical harmonic degree ($l$) in (c-h) for the annotated cases in (a,b). The representative cases from WN (c,d), RR (e,f), and WR (g,h) regimes correspond to $\widetilde{Ra} = 30, 900,$ and $13000,$ respectively.

# Supplementary Material S2: Forces balance at $E = 10^{-6}$

October 4, 2024

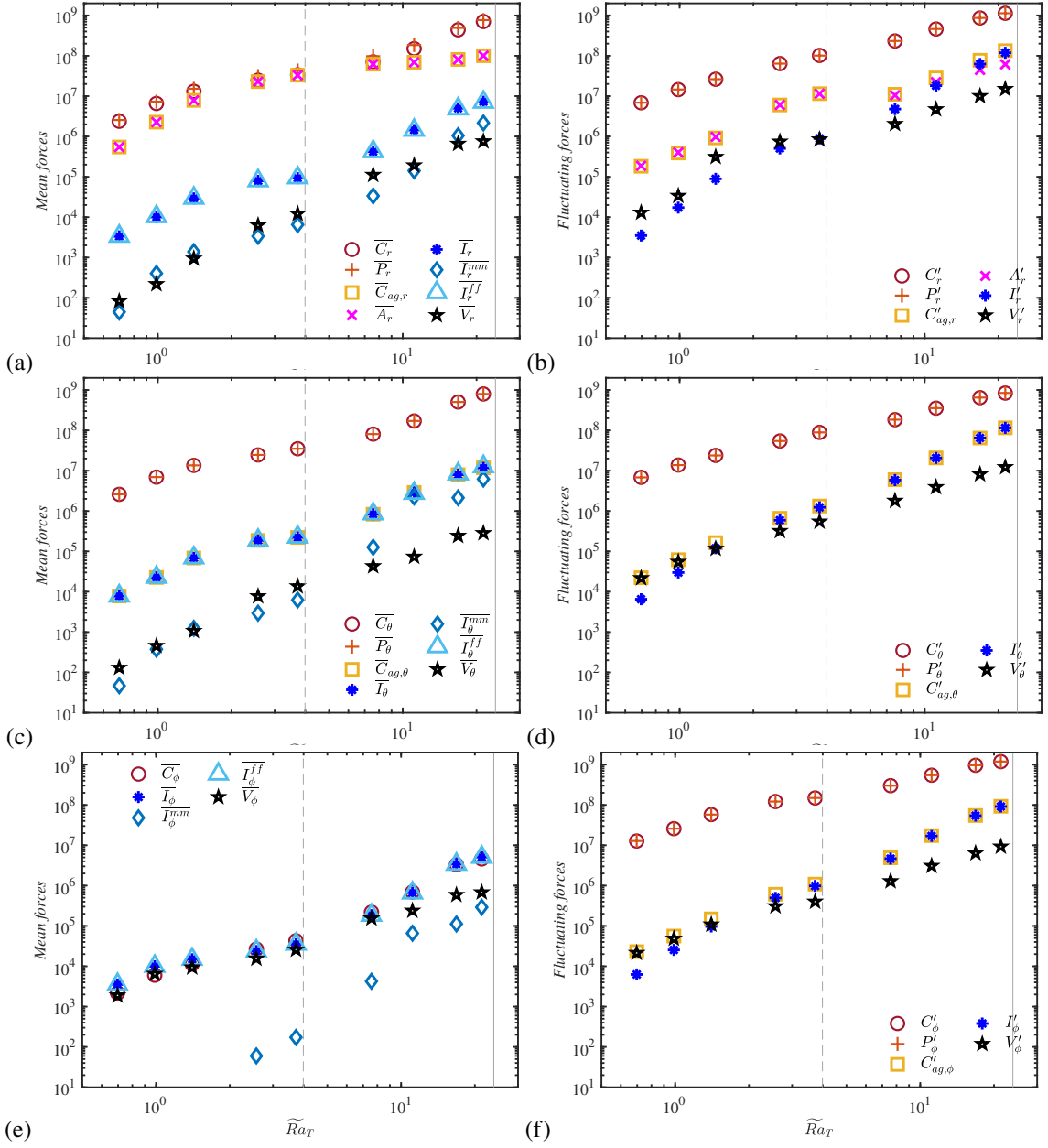

Figure 1: Volume-averaged r.m.s. mean (left column) and fluctuating (right column) force components in $\hat{r}$ (a,b), $\hat{\theta}$ (c,d), and $\hat{\phi}$ (e,f) for $E = 10^{-6}$. The dashed (solid) vertical line represents the thermal forcing where a transition from WN to RR (RR to WR) regime happens according to the scaling predictions of Long *et al.* (2020).

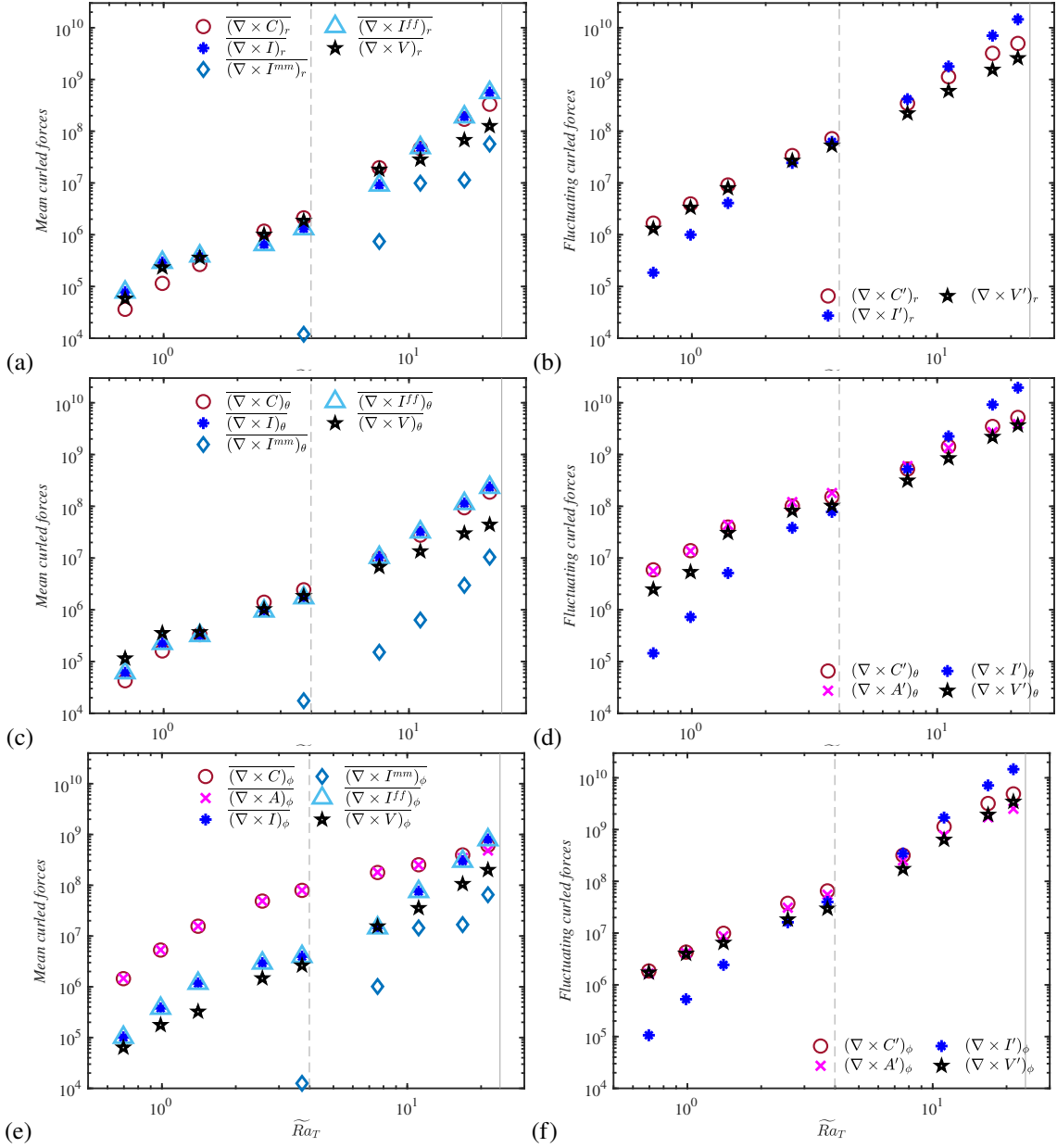

Figure 2: Volume-averaged r.m.s. mean (left column) and fluctuating (right column) curled force components in $\hat{r}$ (a,b), $\hat{\theta}$ (c,d), and $\hat{\phi}$ (e,f) for $E = 10^{-6}$. The dashed (solid) vertical lines represent the thermal forcing where a transition from WN to RR (RR to WR) regime happens according to the scaling predictions of Long *et al.* (2020).

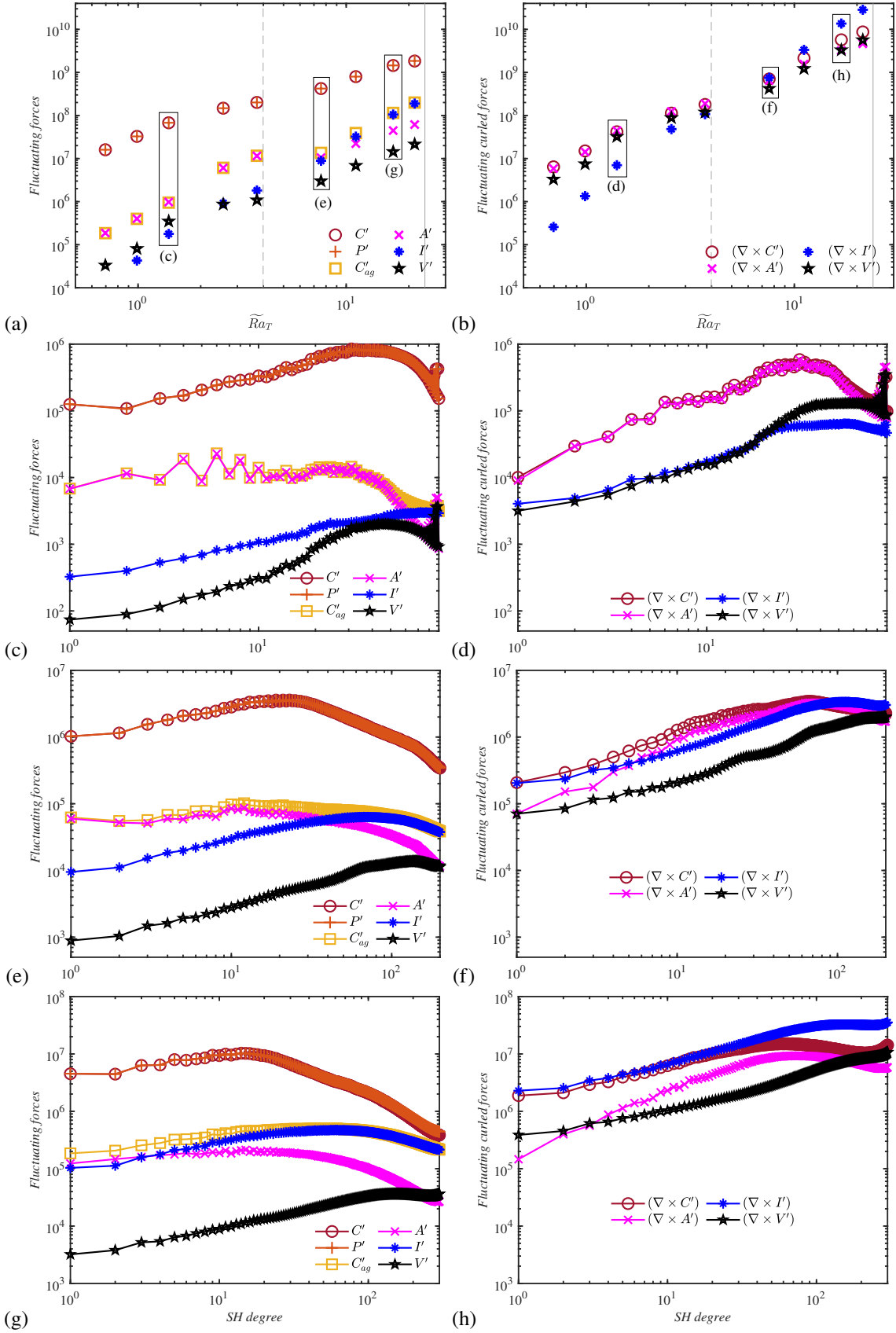

Figure 3: Volume-averaged rms fluctuating forces (left row) and their curls (right row) at $E = 10^{-6}$ as a function of thermal forcing $\widetilde{Ra}_T$ in (a,b) and spherical harmonic degree ($l$) in (c-h) for the annotated cases in (a,b)

# Supplementary data S3: Total mean force magnitudes

October 4, 2024

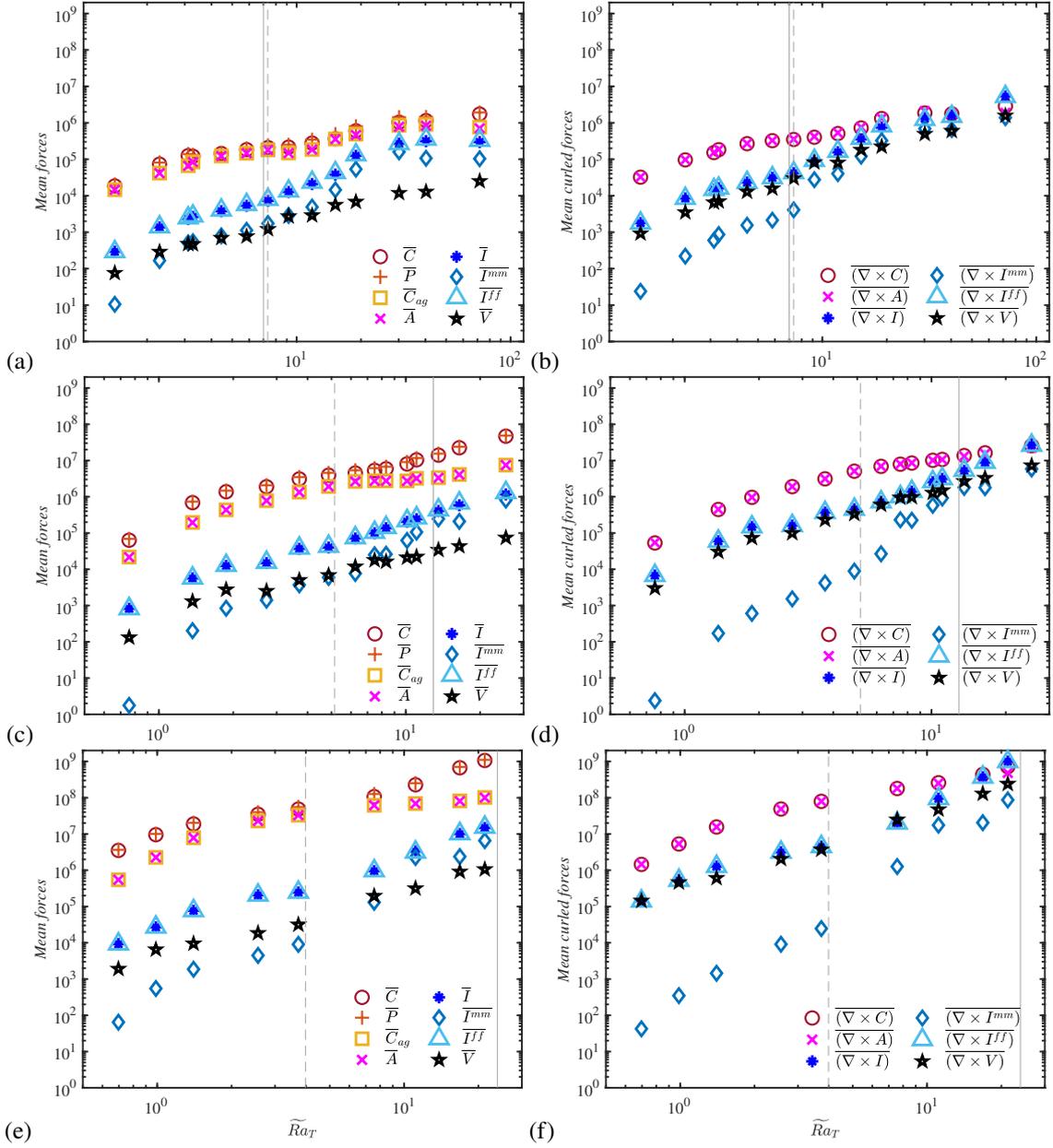

Figure 1: Volume-averaged r.m.s. mean force (left column) and curled force (right column) magnitudes for $E = 10^{-4}$ (a,b), $E = 10^{-5}$ (c,d), and $E = 10^{-6}$ (e,f). The dashed (solid) vertical line represents the thermal forcing where a transition from WN to RR (RR to WR) regime happens according to the scaling predictions of Long *et al.* (2020).